\documentclass[11pt]{article}
\usepackage[margin=1in]{geometry}
\usepackage[usenames,dvipsnames,svgnames,table]{xcolor} % usenames : 16 basic colors,  dvipsnames : other 68 colors, svgnames : another 150 colors. It is used instead of \usepackage{color}. Can find descriptions for color in xcolor package. 

\usepackage{longtable}
\usepackage{hhline}
\usepackage{float}

\usepackage{algorithm}
\usepackage{algorithmic}
\floatname{algorithm}{Algorithm}

\usepackage{setspace}
\usepackage{xspace}
\usepackage{enumitem} % for \begin{description}
\usepackage{rotfloat} % For [H] option in sidewaystable
\usepackage{graphicx}
\usepackage{amssymb, amsmath, amsthm}
\usepackage{verbatim}
\usepackage{natbib}
\usepackage{authblk}
\usepackage{multirow}
\usepackage{subcaption} % for \begin{subtable}
\usepackage{array}
\usepackage{hyperref}
\newcolumntype{L}{>{\centering\arraybackslash}m{0.1\linewidth}}

\newtheorem{theorem}{Theorem}[section]
\newtheorem{lemma}[theorem]{Lemma}
\newtheorem{proposition}[theorem]{Proposition}
\newtheorem{corollary}[theorem]{Corollary}

\newtheorem{definition}{Definition}
\newtheorem{assumption}[theorem]{Assumption}

\newcommand{\Frechet}{Fr\'{e}chet\xspace}
\newcommand{\Poincare}{Poincar\'{e}\xspace}

\DeclareMathOperator*{\argmin}{argmin\xspace}
\DeclareMathOperator*{\argmax}{argmax\xspace}

\newcommand{\bbB}{\mathbb{B}}
\newcommand{\bbH}{\mathbb{H}}
\newcommand{\bbR}{\mathbb{R}}

\usepackage{tikz}
\usepackage{tikz-3dplot}
\usetikzlibrary{arrows.meta,calc,decorations.pathreplacing,positioning}

\begin{document}
	
\title{Finite Mixture Modeling with Riemannian Gaussian Distributions on Hyperbolic Space}
\author[1,2]{Kisung You}
%\author[1,2]{Mystic Cat}
\affil[1]{Department of Mathematics, Baruch College}
\affil[2]{Department of Mathematics, The Graduate Center, City University of New York}
\date{}

\maketitle
\begin{abstract}
Hyperbolic space is increasingly used for hierarchical, tree-like, and network-structured data, but likelihood-based density modeling on hyperbolic space remains relatively limited. This paper develops finite mixture modeling with isotropic Riemannian Gaussian distributions on hyperbolic space under the hyperboloid model. We derive the density, radial normalizing constant, and a finite-sum representation involving the complementary error function.  We then formulate weighted maximum likelihood estimation, which is the fundamental subproblem in mixture fitting: the location estimator is the weighted \Frechet mean, while the inverse-scale estimator is obtained from a one-dimensional strictly convex profile problem. For finite mixtures, we derive exact EM and generalized EM algorithms. The generalized version replaces exact barycenter solves with truncated hyperbolic majorization--minimization updates. We establish existence and uniqueness of the weighted single-component estimator, singularity of the unrestricted mixture likelihood, existence of a constrained mixture estimator, and monotonicity properties of the EM-type algorithms. Simulations show accurate weighted estimation, reliable mixture recovery, effective model selection, and substantial computational savings from generalized EM. Real network examples based on hyperbolic embeddings illustrate the method as an exploratory likelihood-based clustering tool for non-Euclidean data.
\end{abstract}

%%%%%%%%%%%%%%%%%%%%%%%%%%%%%%%%%%%%%%%%%%%%%%%%%%%%%%%%%%%%%%%%%%%%%%%%%%%%%%%%%
\section{Introduction}

A significant trajectory in statistical learning has been the expansion of the class of spaces on which data are modeled. While classical methodology is largely organized around Euclidean geometry, that framework is often inadequate when observations live on constrained or curved domains. A well-known example is directional statistics \citep{mardia_2000_DirectionalStatistics}, which studies data such as orientations, rotations, axes, and subspaces. In such settings, observations no longer obey the usual algebra of vector spaces, and one instead needs models and algorithms that respect the geometry of the sample space. This perspective is central to statistics on manifolds \citep{bhattacharya_2012_NonparametricInferenceManifolds, patrangenaru_2016_NonparametricStatisticsManifolds}, where tools from differential and Riemannian geometry are used to formulate coherent notions of inference, optimization, and uncertainty quantification for non-Euclidean data.

This paradigm is also valuable when the shape of the data itself suggests an underlying geometry. In particular, hyperbolic geometry has become a natural modeling language for hierarchical, tree-like, and exponentially expanding data because negatively curved spaces expand exponentially with radius, mirroring the combinatorial growth of nodes across successive levels of a hierarchy. This geometric compatibility helps explain why trees admit low-distortion embeddings in hyperbolic space and why hyperbolic representations have proved effective for taxonomies, symbolic hierarchies, and related structured data \citep{sarkar_2012_LowDistortionDelaunay, nickel_2017_PoincareEmbeddingsLearning}. The same intuition also underlies hyperbolic models of complex networks, where hierarchy, heterogeneity, and effective tree-likeness are linked to latent hyperbolic geometry \citep{cunningham_2017_NavigabilityRandomGeometric}. Beyond representation, recent work has demonstrated the value of hyperbolic geometry within fully statistical frameworks. For example, \cite{liu_2024_BayesianHyperbolicMultidimensional} proposed a Bayesian multidimensional scaling approach in hyperbolic space, showing that such models can capture hierarchical dependence while enabling regularization and uncertainty quantification. Likewise, \cite{li_2026_HyperbolicNetworkLatent} incorporated a learnable curvature parameter into hyperbolic latent-space models for network analysis, illustrating that curvature itself can play a statistically meaningful role in model specification.

Despite this progress, core inferential methodology on hyperbolic space remains comparatively underdeveloped relative to the much larger literature on hyperbolic representation learning. In particular, basic tasks such as parametric density estimation, likelihood-based clustering, and finite mixture modeling directly on hyperbolic space are still much less developed than their Euclidean or spherical counterparts. Even from a parametric perspective, the systematic study of probability distributions on hyperbolic space is relatively recent. A particularly important development is the work of \cite{said_2018_GaussianDistributionsRiemannian}, who introduced Gaussian distributions on Riemannian symmetric spaces and studied their statistical structure, including mixture-based procedures. More recently, \cite{chen_2024_RiemannianRadialDistributions} placed such models in a broader class of Riemannian radial distributions and established a general theory for parameter estimation on symmetric spaces. These works suggest that hyperbolic geometry is not merely a convenient embedding device, but can serve as a principled domain for likelihood-based statistical modeling.

In this paper, we study a fundamental problem in this direction: finite mixture modeling on hyperbolic space using Riemannian Gaussian distributions. The Riemannian Gaussian may be viewed as the geodesic analogue of the isotropic Gaussian, in the sense that its density depends only on the squared geodesic distance from a location parameter. On symmetric spaces, this model enjoys an especially convenient structure in that the normalizing constant is independent of the location parameter, and the location parameter itself is characterized by a Riemannian barycenter \citep{said_2018_GaussianDistributionsRiemannian, chen_2024_RiemannianRadialDistributions}. These features make the Riemannian Gaussian a natural building block for statistical modeling on hyperbolic space. At the same time, they do not by themselves resolve the main computational difficulty. In finite mixtures, every component-wise M-step requires the repeated solution of weighted \Frechet mean problems \citep{frechet_1948_ElementsAleatoiresNature}, and the unrestricted mixture likelihood inherits the usual singularity associated with scale collapse.

Our contribution is to develop a complete mixture-modeling framework for Riemannian Gaussian distributions on hyperbolic space. First, we formulate finite Riemannian Gaussian mixtures on the hyperboloid and show that weighted single-component estimation is the fundamental primitive underlying the mixture M-step. Second, we develop practical computational procedures for the model, including an exact expectation--maximization (EM) algorithm \citep{dempster_1977_MaximumLikelihoodIncomplete}, a generalized EM variant based on truncated inner updates, and a hyperbolic majorization--minimization routine for the weighted barycenter subproblem. Third, we establish the main theoretical guarantees needed to support likelihood-based computation including existence and uniqueness of the weighted single-component estimator under explicit conditions, singularity of the unrestricted mixture likelihood, existence of the constrained mixture estimator, and monotonicity and convergence properties of the proposed EM-type algorithms. In this sense, the present work aims to place hyperbolic mixture modeling on a footing comparable to more classical model-based clustering methodology, while fully respecting the negative-curvature geometry of the underlying space.

The remainder of the paper is organized as follows. 
Section~\ref{sec:background} reviews the hyperboloid model, the Poincar\'e representation, geodesic convexity, and barycenters. 
Section~\ref{sec:model} develops the Riemannian Gaussian model and the weighted maximum likelihood problem. 
Section~\ref{sec:computation} introduces finite mixtures and the exact and generalized EM algorithms. 
Section~\ref{sec:theory} establishes the main existence, uniqueness, and monotonicity results. 
Section~\ref{sec:experiment} presents simulation studies and real-data examples, and Section~\ref{sec:conclusion} concludes with discussion on future directions.

%%%%%%%%%%%%%%%%%%%%%%%%%%%%%%%%%%%%%%%%%%%%%%%%%%%%%%%%%%%%%%%%%%%%%%%%%%%%%%%%%
\section{Background}\label{sec:background}

We collect the geometric notation and analytic conventions used throughout the paper. Unless stated otherwise, hyperbolic space is taken to have constant sectional curvature of $-1$. Since all computations in this paper are carried out in the Lorentz, or hyperboloid, model, we shall use the same symbol \(\mathbb{H}^d\) to denote both the abstract \(d\)-dimensional hyperbolic space and its realization as the upper sheet of the hyperboloid in \(\mathbb{R}^{d+1}\). This mild abuse of notation is standard and causes little ambiguity, because the quantities that enter our model, such as distance, geodesics, barycenters, and probability densities, are intrinsic even when written in ambient coordinates.

\subsection{Hyperbolic manifold and its geometry}\label{subsec:background_geometry}

We work in \(\mathbb{R}^{d+1}\) equipped with the Lorentzian bilinear form
\begin{equation}\label{eq:bg_lorentz}
\langle x,y\rangle_L
=
\sum_{j=1}^d x_j y_j - x_{d+1}y_{d+1},
\qquad
x=(x_1,\ldots,x_d,x_{d+1}),\ y=(y_1,\ldots,y_d,y_{d+1}).
\end{equation}
The \(d\)-dimensional hyperbolic space is then realized as the upper sheet of the two-sheeted hyperboloid
\begin{equation}\label{eq:bg_hyperboloid}
\mathbb{H}^d
=
\left\{
x\in\mathbb{R}^{d+1}:\ \langle x,x\rangle_L=-1,\ x_{d+1}>0
\right\}.
\end{equation}
Under this convention, the distinguished origin is denoted by $o=(0,\ldots,0,1)$.  The point \(o\) plays the role of the Euclidean origin in the sense that geodesic polar coordinates and many later formulas simplify when they are centered at \(o\).

\begin{figure}[H]
\centering

\tdplotsetmaincoords{70}{110}
\begin{tikzpicture}[tdplot_main_coords, scale=1.8, >=Latex]

% -------------------------------------------------
% Axes
% -------------------------------------------------
\draw[->, thick] (0,0,0) -- (2.8,0,0) node[below left] {$x$};
\draw[->, thick] (0,0,0) -- (0,2.4,0) node[right] {$y$};
\draw[->, thick] (0,0,-1.25) -- (0,0,3.6) node[above] {$z$};

% -------------------------------------------------
% Hyperboloid
% -------------------------------------------------
\foreach \u in {0.3,0.6,0.9,1.2,1.5,1.8}{
  \draw[blue!70, line width=0.3pt, samples=100, smooth, variable=\t, domain=0:360]
    plot ({sinh(\u)*cos(\t)},
          {sinh(\u)*sin(\t)},
          {cosh(\u)});
}

\foreach \v in {0,30,...,330}{
  \draw[blue!60, line width=0.15pt, samples=70, smooth, variable=\u, domain=0.1:1.9]
    plot ({sinh(\u)*cos(\v)},
          {sinh(\u)*sin(\v)},
          {cosh(\u)});
}

% -------------------------------------------------
% Poincare disk on the plane z=0
% -------------------------------------------------
\foreach \u in {0.3,0.6,0.9,1.2,1.5,1.8}{
    \draw[blue!60, line width=0.3pt, samples=140, smooth, variable=\t, domain=0:360]
    plot ({\u/1.8*cos(\t)}, {\u/1.8*sin(\t)}, 0);
}

\node[below right=2pt] at (0,0,0) {$0$};
%\node at (0.25,-1.15,0) {$\mathbb{B}^2$};

% -------------------------------------------------
% Distinguished points
% -------------------------------------------------
\coordinate (O) at (0,0,1);
\coordinate (S) at (0,0,-1);

% -------------------------------------------------
% Two points p and q on the hyperboloid
% -------------------------------------------------
\pgfmathsetmacro{\tp}{-1.8}
\pgfmathsetmacro{\tq}{1.8}

\coordinate (P) at ({sinh(\tp)},0,{cosh(\tp)});
\coordinate (Q) at ({sinh(\tq)},0,{cosh(\tq)});

% -------------------------------------------------
% Projections to the Poincare disk:
% pi(x,y,z) = (x/(1+z), y/(1+z), 0)
% -------------------------------------------------
\pgfmathsetmacro{\xp}{sinh(\tp)/(1+cosh(\tp))}
\pgfmathsetmacro{\xq}{sinh(\tq)/(1+cosh(\tq))}

\coordinate (PP) at (\xp,0,0);
\coordinate (QQ) at (\xq,0,0);

% -------------------------------------------------
% Geodesic on the hyperboloid
% -------------------------------------------------
\draw[red!80!black, line width=1.4pt, samples=120, smooth, variable=\t, domain=\tp:\tq]
  plot ({sinh(\t)},0,{cosh(\t)});

% -------------------------------------------------
% Euclidean chord in ambient space
% -------------------------------------------------
\draw[orange!85!black, dashed, line width=1.2pt] (P) -- (Q);

% -------------------------------------------------
% Projection lines from south pole (0,0,-1)
% These pass through pi(p) and pi(q), respectively.
% -------------------------------------------------
\draw[gray!75, densely dotted, line width=1.0pt] (S) -- (P);
\draw[gray!75, densely dotted, line width=1.0pt] (S) -- (Q);

% -------------------------------------------------
% Mark the distinguished points
% -------------------------------------------------
\fill (O) circle (0.8pt);
\node[below=2pt, xshift=30pt] at (O) {$o=(0,0,1)$};

\fill (S) circle (0.8pt);
\node[right=2pt] at (S) {$(0,0,-1)$};

\fill (P) circle (0.9pt);
\fill (Q) circle (0.9pt);

\node[above left=2pt] at (P) {$p$};
\node[above left=5pt] at (Q) {$q$};

\fill (PP) circle (0.8pt);
\fill (QQ) circle (0.8pt);

\node[right=3pt] at (PP) {$\pi(p)$};
\node[left=2pt] at (QQ) {$\pi(q)$};

% -------------------------------------------------
% Geodesic in the Poincare disk
% Since p and q lie on the y=0 geodesic in the hyperboloid,
% their images lie on a diameter, so the Poincare geodesic
% is the straight red segment.
% -------------------------------------------------
\draw[red!80!black, line width=1.4pt] (PP) -- (QQ);

\end{tikzpicture}

\caption{
Hyperboloid and Poincar\'e representations of the two-dimensional hyperbolic space. 
The upper sheet of  $\bbH^2$ is shown together with its stereographic projection onto the plane \(z=0\). 
The red curve denotes the hyperbolic geodesic connecting \(p\) and \(q\), the orange dashed segment denotes the ambient Euclidean chord, and the gray dotted lines indicate projection from the south pole \((0,0,-1)\) to the Poincar\'e disk.
}
\label{fig:hyperboloid_poincare_overlay}
\end{figure}
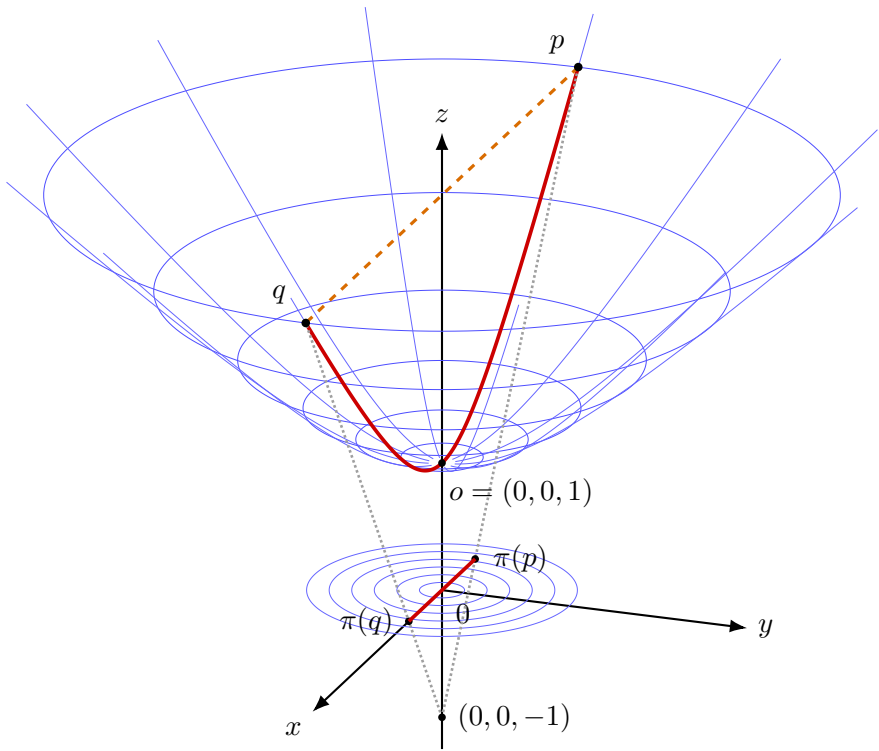

Figure~\ref{fig:hyperboloid_poincare_overlay} illustrates two equivalent realizations of the same hyperbolic space. The upper sheet of the 2-dimensional hyperboloid $\bbH^2 \subset \bbR^3$ is shown together with its stereographic projection from the south pole \((0,0,-1)\) onto the plane \(z=0\), yielding the \Poincare disk $\bbB^2$. The red curve on the hyperboloid is a geodesic connecting two points \(p\) and \(q\), while the orange dashed segment is the chord in the ambient Euclidean space. This contrast emphasizes that geodesics on \(\bbH^d\) are determined by the intrinsic hyperbolic metric, not by straight lines in the surrounding Euclidean space. The \Poincare model is visually intuitive, but the hyperboloid model is computationally preferable here because the distance, exponential map, logarithm map, and tangent-space projection all admit simple formulas with better numerical stability \citep{mishne_2023_NumericalStabilityHyperbolic}. 

% -----------------------------------------------------------------
% Keep your figure here.
% -----------------------------------------------------------------

For any \(\mu\in\mathbb{H}^d\), the tangent space is
\begin{equation}\label{eq:bg_tangent}
T_\mu\mathbb{H}^d
=
\left\{
v\in\mathbb{R}^{d+1}:\ \langle \mu,v\rangle_L=0
\right\}.
\end{equation}
If \(v\in T_\mu\mathbb{H}^d\), then \(\langle v,v\rangle_L>0\) unless \(v=0\), so the Riemannian norm of a tangent vector is
\begin{equation}\label{eq:bg_tangent_norm}
\|v\|_L=\sqrt{\langle v,v\rangle_L}.
\end{equation}
The Lorentz-orthogonal projection of an ambient vector \(u\in\mathbb{R}^{d+1}\) onto \(T_\mu\mathbb{H}^d\) is
\begin{equation}\label{eq:bg_projection}
\operatorname{Proj}_\mu(u)
=
u+\langle \mu,u\rangle_L\,\mu,
\end{equation}
whose validity can be checked by the following equivalence:
\[
\left\langle \mu,\operatorname{Proj}_\mu(u)\right\rangle_L
=
\langle \mu,u\rangle_L+\langle \mu,u\rangle_L\langle \mu,\mu\rangle_L
=
\langle \mu,u\rangle_L-\langle \mu,u\rangle_L
=
0.
\]

For two points $x,y \in \bbH^d$, the geodesic distance is
\begin{equation}\label{eq:bg_distance}
d(x,y)= \operatorname{arcosh}\bigl(-\langle x,y\rangle_L\bigr),
\end{equation}
where \(\operatorname{arcosh}\) denotes the inverse hyperbolic cosine. In particular, since
$\langle o,x\rangle_L=-x_{d+1},$ the distance to $x$ from the origin is simplified as $d(o,x)=\operatorname{arcosh}(x_{d+1})$. This identity explains why the last coordinate naturally plays the role of a radial variable in the hyperboloid model.

Now we describe two operations that connect the tangent space and the manifold. Given a point  \(\mu\in\mathbb{H}^d\) and a tangent vector \(v\in T_\mu\mathbb{H}^d\), the exponential map 
\begin{equation}\label{eq:bg_exp}
\operatorname{Exp}_\mu(v)
=
\cosh(\|v\|_L)\mu
+
\sinh(\|v\|_L)\frac{v}{\|v\|_L},
\end{equation}
with the usual continuous interpretation \(\operatorname{Exp}_\mu(0)=\mu\). Equivalently, the geodesic with initial condition \(\gamma(0)=\mu\) and \(\dot{\gamma}(0)=v\) is
\begin{equation}\label{eq:bg_geodesic}
\gamma(t)=\operatorname{Exp}_\mu(tv)
=
\cosh\!\bigl(t\|v\|_L\bigr)\mu
+
\sinh\!\bigl(t\|v\|_L\bigr)\frac{v}{\|v\|_L}.
\end{equation}
Thus geodesics in the hyperboloid model are obtained by intersecting \(\mathbb{H}^d\) with two-dimensional linear subspaces of \(\mathbb{R}^{d+1}\) passing through the origin.

The logarithm map is the inverse of the exponential map along geodesics. For \(\mu,x\in\mathbb{H}^d\), let
$\alpha(\mu,x):=-\langle \mu,x\rangle_L\ge 1$. Then the logarithm map is defined as 
\begin{equation}\label{eq:bg_log}
\operatorname{Log}_\mu(x)
=
\frac{\operatorname{arcosh}\bigl(\alpha(\mu,x)\bigr)}{\sqrt{\alpha(\mu,x)^2-1}}
\left(x-\alpha(\mu,x)\mu\right).
\end{equation}
At the origin \(o=(0,\ldots,0,1)\), this simplifies to
\begin{equation}\label{eq:bg_log_origin}
\operatorname{Log}_o(x)
=
\frac{\operatorname{arcosh}(x_{d+1})}{\sqrt{x_{d+1}^2-1}}
(x_1,\ldots,x_d,0).
\end{equation}
Later, these explicit formulas for the distance and logarithm map will allow us to write both the likelihood and the weighted barycenter equations in a computationally convenient form.

Although we carry out all computations on the hyperboloid, it is useful to record the corresponding Poincar\'e ball representation. Let $\mathbb{B}^d=\{u\in\mathbb{R}^d:\ \|u\|<1\}$ denote the interior of the unit ball, 
where \(\|\cdot\|\) is the usual Euclidean norm on \(\mathbb{R}^d\). Writing a point on the hyperboloid as \(x=(u,t)\) with \(u\in\mathbb{R}^d\) and \(t=x_{d+1}\), the stereographic projection from the south pole \((0,\ldots,0,-1)\) onto the plane \(x_{d+1}=0\) is
\begin{equation}\label{eq:bg_stereographic}
\pi(u,t)=\frac{u}{1+t}\in\mathbb{B}^d.
\end{equation}
Its inverse is
\begin{equation}\label{eq:bg_stereographic_inverse}
\pi^{-1}(y)
=
\left(
\frac{2y}{1-\|y\|^2},
\frac{1+\|y\|^2}{1-\|y\|^2}
\right),
\qquad y\in\mathbb{B}^d.
\end{equation}
The two models are isometric, so they encode exactly the same hyperbolic geometry. In the Poincar\'e ball, geodesics appear as Euclidean diameters or circular arcs orthogonal to the unit sphere, whereas in the hyperboloid model they appear as intersections with linear subspaces through the origin. In particular, the projected red curve in Figure~\ref{fig:hyperboloid_poincare_overlay} is the same hyperbolic geodesic shown in a different coordinate system.

\subsection{Geodesic convexity and barycenters}\label{subsec:background_barycenter}

A key reason hyperbolic space is statistically tractable is that it is a Hadamard manifold, i.e., it is complete, simply connected, and has nonpositive sectional curvature. As a consequence, any two points in \(\mathbb{H}^d\) are joined by a unique geodesic, and optimization problems built from squared geodesic distances enjoy strong convexity properties. This is precisely the feature that makes barycenters, maximum likelihood estimation, and mixture-model updates well behaved in the present setting.

For a geodesic \(\gamma:[0,1]\to\mathbb{H}^d\) and a fixed point \(x\in\mathbb{H}^d\), the squared distance satisfies the Hadamard-space inequality
\begin{equation}\label{eq:bg_strong_convexity}
d^2\!\bigl(x,\gamma(t)\bigr)
\le
(1-t)d^2\!\bigl(x,\gamma(0)\bigr)
+
td^2\!\bigl(x,\gamma(1)\bigr)
-
t(1-t)d^2\!\bigl(\gamma(0),\gamma(1)\bigr).
\end{equation}
In particular, \(d^2(x,\cdot)\) is geodesically convex, and finite weighted sums of such functions are in fact strongly geodesically convex. This fact will later guarantee the uniqueness of the location estimator in the Riemannian Gaussian model.

Let \(x_1,\ldots,x_n\in\mathbb{H}^d\) and let \(w_1,\ldots,w_n\ge 0\) be nonnegative weights, not all zero. Write
\begin{equation}\label{eq:bg_weight_sum}
W=\sum_{i=1}^n w_i>0.
\end{equation}
The weighted Fr\'echet functional is
\begin{equation}\label{eq:bg_weighted_frechet}
F_w(\mu)
=
\sum_{i=1}^n w_i\, d^2(x_i,\mu),
\qquad \mu\in\mathbb{H}^d.
\end{equation}
Its unique minimizer
\begin{equation}\label{eq:bg_weighted_barycenter}
\mu_w^\star
=
\argmin_{\mu\in\mathbb{H}^d} F_w(\mu)
\end{equation}
is called the weighted Fr\'echet mean, or weighted barycenter, of the sample \citep{afsari_2011_Riemannian$L_p$Center}. In the single-component Riemannian Gaussian model, it is exactly the maximum likelihood estimator of the location parameter. In finite mixtures, it becomes the component-wise location update in the M-step of the EM algorithm.

It is also useful to view the weighted sample as a discrete probability measure
\begin{equation}\label{eq:bg_weighted_measure}
\nu_w
=
\frac{1}{W}\sum_{i=1}^n w_i \delta_{x_i},
\end{equation}
where \(\delta_{x_i}\) denotes the point mass at \(x_i\). Then \(F_w(\mu)=W\int d^2(x,\mu)\,\nu_w(dx)\), so the weighted barycenter is the Fr\'echet mean of \(\nu_w\). In this sense, the finite-sample barycenter problem and the population barycenter problem are the same optimization problem at different levels of abstraction.

For later computational developments, we record the first-order characterization of the weighted barycenter. The Riemannian gradient of \(F_w\) is
\begin{equation}\label{eq:bg_barycenter_gradient}
\nabla F_w(\mu)
=
-2\sum_{i=1}^n w_i\,\operatorname{Log}_\mu(x_i),
\end{equation}
and therefore the unique minimizer \(\mu_w^\star\) satisfies the score equation
\begin{equation}\label{eq:bg_barycenter_score}
\sum_{i=1}^n w_i\,\operatorname{Log}_{\mu_w^\star}(x_i)=0.
\end{equation}
This equation is the hyperbolic analogue of the Euclidean identity that the centered sample mean has zero average residual. Its weighted form will later reappear as the component-wise likelihood equation in our mixture model.

Finally, we introduce the geodesic convex hull notation that will be used to define a constrained parameter space for finite mixtures. For a subset \(A\subset\mathbb{H}^d\), its geodesic convex hull, denoted \(\operatorname{conv}_g(A)\), is the smallest geodesically convex set containing \(A\). Since closed metric balls in \(\mathbb{H}^d\) are geodesically convex, the closed geodesic convex hull of finitely many points is compact. This simple observation will later allow us to constrain location parameters to a natural compact subset of hyperbolic space without sacrificing statistical interpretability.

\subsection{Special functions}\label{subsec:background_special}

The normalizing constants of Riemannian Gaussian distributions on hyperbolic space involve a few classical special functions. We record them here once and for all so that later formulas can be written without interruption.

First, for \(u\ge 1\), the inverse hyperbolic cosine is
\begin{equation}\label{eq:bg_arcosh}
\operatorname{arcosh}(u)
=
\log\!\left(u+\sqrt{u^2-1}\right).
\end{equation}
This function appears throughout hyperbolic geometry because the Lorentzian distance formula \eqref{eq:bg_distance} is written in terms of \(\operatorname{arcosh}\).

Second, we define the error function and complementary error function by
\begin{equation*}
\operatorname{erf}(z)
=
\frac{2}{\sqrt{\pi}}\int_0^z e^{-t^2}\,dt,\quad \operatorname{erfc}(z)
=
1-\operatorname{erf}(z)
=
\frac{2}{\sqrt{\pi}}\int_z^\infty e^{-t^2}\,dt.
\end{equation*}
The function \(\operatorname{erfc}\) will arise naturally when we derive an exact finite-sum representation of the normalizing constant for the Riemannian Gaussian distribution. Concretely, after expanding a power of \(\sinh(r)\) and completing the square in Gaussian tail integrals over \([0,\infty)\), the remaining terms can be written in closed form using \(\operatorname{erfc}\). Since this function is less familiar in statistics than, say, the Gamma function, it is helpful to define it explicitly at the outset.

Finally, we shall use
\begin{equation}\label{eq:bg_omega}
\omega_{d-1}
=
\frac{2\pi^{d/2}}{\Gamma(d/2)}
\end{equation}
to denote the surface area of the Euclidean unit sphere \(S^{d-1}\subset\mathbb{R}^d\), where \(\Gamma\) is the Gamma function. This constant appears when hyperbolic volume is written in geodesic polar coordinates. More specifically, if \(r\ge 0\) denotes the hyperbolic distance from the origin and \(u\in S^{d-1}\) is a direction, then the hyperbolic volume element takes the form
\begin{equation}\label{eq:bg_volume_element}
d\operatorname{vol}
=
\sinh^{d-1}(r)\,dr\,d\omega(u),
\end{equation}
where \(d\omega(u)\) is the surface measure on \(S^{d-1}\). The factor \(\sinh^{d-1}(r)\) reflects the exponential volume growth characteristic of negative curvature, and it is the geometric reason why hyperbolic Gaussian normalizing constants differ fundamentally from their Euclidean counterparts.

%%%%%%%%%%%%%%%%%%%%%%%%%%%%%%%%%%%%%%%%%%%%%%%%%%%%%%%%%%%%%%%%%%%%%%%%%%%%%%%%%
%%%%%%%%%%%%%%%%%%%%%%%%%%%%%%%%%%%%%%%%%%%%%%%%%%%%%%%%%%%%%%%%%%%%%%%%%%%%%%%%%
\section{Riemannian Gaussian distributions and parameter estimation}\label{sec:model}

We now introduce the parametric family that will serve as the basic building block throughout the paper. Following \cite{said_2018_GaussianDistributionsRiemannian}, we consider the isotropic {Riemannian Gaussian distribution} on hyperbolic space, also referred to in parts of the literature as the {Riemannian normal}. The model is defined intrinsically through the squared geodesic distance, and may therefore be viewed as the natural continuation of the Euclidean isotropic Gaussian to a negatively curved space. From the standpoint of likelihood-based inference, this family is especially attractive on \(\mathbb{H}^d\). The location parameter is tied to a barycenter problem, while the scale parameter is governed by a one-dimensional profile function.

Although our ultimate goal is finite mixture modeling, the appropriate primitive is not the ordinary i.i.d.\ likelihood but the weighted single-component problem. This is because, in the mixture setting, component-wise M-step updates are obtained by replacing sample counts with posterior responsibilities. For this reason, we develop the weighted formulation from the outset and treat the unweighted case as the special choice \(w_i\equiv 1\).

\subsection{Riemannian Gaussian distributions and their normalizing constant}\label{subsec:model-definition}

The Riemannian Gaussian on \(\mathbb{H}^d\) is a radial distribution centered at \(\mu\in\mathbb{H}^d\), whose density depends only on the squared geodesic distance \(d^2(x,\mu)\). We parameterize scale through the inverse-scale parameter \(\beta>0\), which will later be convenient for profile likelihood calculations and EM updates.

\begin{definition}[Riemannian Gaussian distribution on \(\mathbb{H}^d\)]\label{def:intrinsic-gaussian}
For \(\mu\in\mathbb{H}^d\) and \(\beta>0\), the isotropic Riemannian Gaussian distribution is the probability measure with density
\begin{equation}\label{eq:intrinsic-gaussian-density}
p(x\mid \mu,\beta)
=
\frac{1}{Z_d(\beta)}
\exp\bigl\{-\beta\, d^2(x,\mu)\bigr\},
\qquad x\in\mathbb{H}^d,
\end{equation}
with respect to the Riemannian volume measure on \(\mathbb{H}^d\). Equivalently, if \(\sigma^2=(2\beta)^{-1}\), then
\begin{equation}\label{eq:intrinsic-gaussian-sigma}
p(x\mid \mu,\sigma)
=
\frac{1}{Z_d\bigl((2\sigma^2)^{-1}\bigr)}
\exp\!\left\{-\frac{d^2(x,\mu)}{2\sigma^2}\right\}.
\end{equation}
\end{definition}

The distribution in \eqref{eq:intrinsic-gaussian-density} is intrinsic in the sense that it depends only on the geodesic distance on \(\mathbb{H}^d\). Because hyperbolic space is homogeneous, the normalizing constant does not depend on the location parameter \(\mu\). Indeed, any point of \(\mathbb{H}^d\) can be moved to the origin \(o\) by an isometry, and both the geodesic distance and the volume measure are invariant under such transformations. Consequently, it suffices to compute the normalizing constant in geodesic polar coordinates centered at \(o\).

Using the notation of \eqref{eq:bg_volume_element}, the hyperbolic volume form in radial coordinates is
\[
d\operatorname{vol}
=
\sinh^{d-1}(r)\,dr\,d\omega(u),
\qquad r\ge 0,\quad u\in S^{d-1},
\]
where \(\omega_{d-1}=2\pi^{d/2}/\Gamma(d/2)\) is the surface area of the Euclidean unit sphere \(S^{d-1}\subset\mathbb{R}^d\). Therefore the normalizing constant is
\begin{equation}\label{eq:normalizer-polar}
Z_d(\beta)
=
\omega_{d-1}\int_0^\infty e^{-\beta r^2}\sinh^{d-1}(r)\,dr.
\end{equation}
The factor \(\sinh^{d-1}(r)\) is the hyperbolic equivalent of the Euclidean radial Jacobian \(r^{d-1}\), and it reflects the exponential growth of volume in negative curvature. This observation is the main geometric reason why the hyperbolic Gaussian normalizer differs from its Euclidean counterpart.

It is also useful to record the induced radial distribution. If \(R=d(o,X)\) for \(X\sim p(\cdot\mid o,\beta)\), then the density of \(R\) on \([0,\infty)\) is proportional to
\begin{equation}\label{eq:radial-density}
r\longmapsto e^{-\beta r^2}\sinh^{d-1}(r).
\end{equation}
Thus the Riemannian Gaussian is determined by a Gaussian-like radial decay combined with a hyperbolic volume correction.

The integral representation \eqref{eq:normalizer-polar} is already enough for likelihood evaluation and numerical quadrature. Nevertheless, for later asymptotic and computational purposes, it is convenient to have an exact finite-sum representation. The next proposition provides such a formula.

\begin{proposition}[Finite-sum representation of the normalizing constant]\label{prop:normalizer-closed-form}
Let
\begin{equation}\label{eq:cj-def}
c_j=2j-d+1,
\qquad j=0,1,\ldots,d-1.
\end{equation}
Then
\begin{equation}\label{eq:normalizer-closed-form}
Z_d(\beta)
=
\frac{\omega_{d-1}\sqrt{\pi}}{2^d\sqrt{\beta}}
\sum_{j=0}^{d-1}
(-1)^{d-1-j}\binom{d-1}{j}
\exp\!\left(\frac{c_j^2}{4\beta}\right)
\operatorname{erfc}\!\left(-\frac{c_j}{2\sqrt{\beta}}\right).
\end{equation}
\end{proposition}

\begin{proof}
Starting from the identity
\[
\sinh r=\frac{e^r-e^{-r}}{2},
\]
we expand
\[
\sinh^{d-1}(r)
=
\left(\frac{e^r-e^{-r}}{2}\right)^{d-1}
=
2^{1-d}\sum_{j=0}^{d-1}
(-1)^{d-1-j}\binom{d-1}{j}e^{(2j-d+1)r}
=
2^{1-d}\sum_{j=0}^{d-1}
(-1)^{d-1-j}\binom{d-1}{j}e^{c_j r}.
\]
Substituting this expression into \eqref{eq:normalizer-polar} gives
\[
Z_d(\beta)
=
\frac{\omega_{d-1}}{2^{d-1}}
\sum_{j=0}^{d-1}
(-1)^{d-1-j}\binom{d-1}{j}
\int_0^\infty e^{-\beta r^2+c_j r}\,dr.
\]
Thus it remains to compute
\[
I(c,\beta):=\int_0^\infty e^{-\beta r^2+cr}\,dr.
\]
Completing the square,
\[
-\beta r^2+cr
=
-\beta\left(r-\frac{c}{2\beta}\right)^2+\frac{c^2}{4\beta},
\]
so
\[
I(c,\beta)
=
\exp\!\left(\frac{c^2}{4\beta}\right)
\int_0^\infty
\exp\!\left[-\beta\left(r-\frac{c}{2\beta}\right)^2\right]dr.
\]
Now set
\[
\nu=\sqrt{\beta}\left(r-\frac{c}{2\beta}\right),
\qquad
dr=\frac{d\nu}{\sqrt{\beta}},
\qquad
r=0\iff \nu=-\frac{c}{2\sqrt{\beta}}.
\]
Then
\[
I(c,\beta)
=
\frac{1}{\sqrt{\beta}}
\exp\!\left(\frac{c^2}{4\beta}\right)
\int_{-c/(2\sqrt{\beta})}^{\infty} e^{-\nu^2}\,d\nu.
\]
By the definition of the complementary error function, we have
\[
\int_a^\infty e^{-\nu^2}\,d\nu
=
\frac{\sqrt{\pi}}{2}\operatorname{erfc}(a),
\]
and therefore
\[
I(c,\beta)
=
\frac{\sqrt{\pi}}{2\sqrt{\beta}}
\exp\!\left(\frac{c^2}{4\beta}\right)
\operatorname{erfc}\!\left(-\frac{c}{2\sqrt{\beta}}\right).
\]
Substituting this back into the finite sum proves \eqref{eq:normalizer-closed-form}.
\end{proof}

The two representations \eqref{eq:normalizer-polar} and \eqref{eq:normalizer-closed-form} play complementary roles. The radial integral is often the most stable form for numerical quadrature, whereas the finite-sum expression is useful for asymptotic calculations and for precomputing the normalizer on a grid of \(\beta\)-values.

\subsection{The log-normalizer and radial moment structure}\label{subsec:lognormalizer}

For likelihood-based estimation, it is convenient to work with the log-normalizer
\begin{equation}\label{eq:log-normalizer}
A_d(\beta):=\log Z_d(\beta).
\end{equation}
The derivatives of \(A_d\) have a particularly transparent interpretation in terms of the radial variable \(R\) from \eqref{eq:radial-density}. This observation shows that scale estimation reduces to a one-dimensional strictly convex problem.

\begin{proposition}[Derivatives of the log-normalizer]\label{prop:Ad-derivatives}
Let \(R\) denote a nonnegative random variable with density proportional to
\[
r\longmapsto e^{-\beta r^2}\sinh^{d-1}(r),
\qquad r\ge 0.
\]
Then
\begin{align}
A_d'(\beta)
&=
-\frac{\int_0^\infty r^2 e^{-\beta r^2}\sinh^{d-1}(r)\,dr}
{\int_0^\infty e^{-\beta r^2}\sinh^{d-1}(r)\,dr}
=
-\mathbb{E}_\beta[R^2],\label{eq:Adprime}
\\
A_d''(\beta)
&=
\operatorname{Var}_\beta(R^2)>0.\label{eq:Addoubleprime}
\end{align}
Consequently, \(A_d\) is strictly convex on \((0,\infty)\), and the function
\begin{equation}\label{eq:m2d-def}
m_{2,d}(\beta):=-A_d'(\beta)=\mathbb{E}_\beta[R^2]
\end{equation}
is strictly decreasing.
\end{proposition}

\begin{proof}
Differentiating \eqref{eq:normalizer-polar} under the integral sign gives
\[
Z_d'(\beta)
=
-\omega_{d-1}\int_0^\infty r^2 e^{-\beta r^2}\sinh^{d-1}(r)\,dr.
\]
Dividing by \(Z_d(\beta)\) yields
\[
A_d'(\beta)=\frac{Z_d'(\beta)}{Z_d(\beta)}=-\mathbb{E}_\beta[R^2].
\]
Differentiating once more and using the standard identity for exponential families gives
\[
A_d''(\beta)
=
\frac{Z_d''(\beta)}{Z_d(\beta)}
-
\left(\frac{Z_d'(\beta)}{Z_d(\beta)}\right)^2
=
\operatorname{Var}_\beta(R^2).
\]
Since the radial distribution is nondegenerate, this variance is strictly positive. The remaining claims follow immediately.
\end{proof}

The next asymptotic result will later explain the scale-collapse phenomenon in finite mixtures.

\begin{proposition}[Large-\(\beta\) asymptotic of the normalizing constant]\label{prop:Zd-asymptotic}
As \(\beta\to\infty\),
\begin{equation}\label{eq:Zd-asymptotic}
Z_d(\beta)\sim \pi^{d/2}\beta^{-d/2}.
\end{equation}
Equivalently,
\begin{equation}\label{eq:Ad-asymptotic}
A_d(\beta)=-\frac{d}{2}\log\beta+O(1),
\qquad \beta\to\infty.
\end{equation}
\end{proposition}

\begin{proof}
For small \(r\), one has \(\sinh r\sim r\). Since the main contribution to \eqref{eq:normalizer-polar} for large \(\beta\) comes from a neighborhood of \(r=0\), the hyperbolic integral is asymptotically equivalent to its Euclidean counterpart:
\[
Z_d(\beta)
\sim
\omega_{d-1}\int_0^\infty e^{-\beta r^2}r^{d-1}\,dr.
\]
The latter integral is classical:
\[
\int_0^\infty e^{-\beta r^2}r^{d-1}\,dr
=
\frac{1}{2}\Gamma\!\left(\frac{d}{2}\right)\beta^{-d/2}.
\]
Multiplying by \(\omega_{d-1}=2\pi^{d/2}/\Gamma(d/2)\) yields
\[
Z_d(\beta)\sim \pi^{d/2}\beta^{-d/2}.
\]
Taking logarithms gives \eqref{eq:Ad-asymptotic}.
\end{proof}

\subsection{Weighted likelihood and maximum likelihood estimation}\label{subsec:weighted-mle}

We now turn to parameter estimation. As noted above, the weighted formulation is the relevant one even if one is ultimately interested in the ordinary sample likelihood, because weighted estimation is exactly what appears in mixture-model M-steps.

Let \(x_1,\ldots,x_n\in\mathbb{H}^d\) and let \(w_1,\ldots,w_n\ge 0\) be nonnegative weights, not all zero. In the remainder of the paper, we write
\begin{equation}\label{eq:weighted-variance}
S_w(\mu)
=
\sum_{i=1}^n w_i\,d^2(x_i,\mu)
\end{equation}
for the weighted Fr\'echet functional introduced in Section~\ref{sec:background}. The weighted log-likelihood under the Riemannian Gaussian model is
\begin{equation}\label{eq:weighted-loglik}
\ell_w(\mu,\beta)
=
\sum_{i=1}^n w_i\log p(x_i\mid\mu,\beta)
=
-\beta S_w(\mu)-W A_d(\beta),
\end{equation}
where \(W=\sum_{i=1}^n w_i\). Therefore the weighted maximum likelihood estimator solves
\begin{equation}\label{eq:weighted-mle-problem}
(\hat{\mu}_w,\hat{\beta}_w)
=
\argmax_{\mu\in\mathbb{H}^d,\ \beta>0}\ell_w(\mu,\beta)
=
\argmin_{\mu\in\mathbb{H}^d,\ \beta>0}
\left\{
\beta S_w(\mu)+W A_d(\beta)
\right\}.
\end{equation}
The structure of \eqref{eq:weighted-loglik} immediately separates location and scale. Indeed,
\begin{equation}\label{eq:location-profile}
\hat{\mu}_w
=
\argmin_{\mu\in\mathbb{H}^d} S_w(\mu),
\end{equation}
and once \(\hat{\mu}_w\) has been obtained, the scale estimate is the solution of the one-dimensional profile problem
\begin{equation}\label{eq:scale-profile}
\hat{\beta}_w
=
\argmin_{\beta>0}
\left\{
\beta S_w(\hat{\mu}_w)+W A_d(\beta)
\right\}.
\end{equation}
Thus the entire estimation problem reduces to two pieces: a weighted barycenter problem for the location parameter and a strictly convex scalar optimization problem for the scale parameter.

The first part has a clear geometric interpretation. Since \(S_w\) is the weighted Fr\'echet functional, the location estimator is exactly the weighted barycenter of the sample. The next proposition records the corresponding score equation.

\begin{proposition}[Weighted score equation for the location parameter]\label{prop:score-equation-location}
The Riemannian gradient of \(S_w\) is
\begin{equation}\label{eq:gradient-Sw}
\nabla S_w(\mu)
=
-2\sum_{i=1}^n w_i\,\operatorname{Log}_\mu(x_i).
\end{equation}
Consequently, any minimizer \(\hat{\mu}_w\) of \(S_w\) satisfies
\begin{equation}\label{eq:score-location}
\sum_{i=1}^n w_i\,\operatorname{Log}_{\hat{\mu}_w}(x_i)=0.
\end{equation}
In ambient coordinates, if
\begin{equation}\label{eq:alpha-i-hat}
\alpha_i=-\langle \hat{\mu}_w,x_i\rangle_L,
\end{equation}
then \eqref{eq:score-location} is equivalent to
\begin{equation}\label{eq:score-location-ambient}
\sum_{i=1}^n
w_i\,
\frac{\operatorname{arcosh}(\alpha_i)}{\sqrt{\alpha_i^2-1}}
\bigl(x_i-\alpha_i\hat{\mu}_w\bigr)
=
0.
\end{equation}
\end{proposition}

\begin{proof}
On a Hadamard manifold, the gradient of the squared-distance function is
\[
\nabla_\mu d^2(x_i,\mu)=-2\,\operatorname{Log}_\mu(x_i).
\]
Summing over \(i\) with weights \(w_i\) gives \eqref{eq:gradient-Sw}. Any minimizer of \(S_w\) must satisfy \(\nabla S_w(\hat{\mu}_w)=0\), yielding \eqref{eq:score-location}. Substituting the explicit logarithm map \eqref{eq:bg_log} then gives \eqref{eq:score-location-ambient}.
\end{proof}

The scale parameter admits a similarly simple characterization. Differentiating the profile objective in \eqref{eq:scale-profile} and using \eqref{eq:m2d-def} yields
\begin{equation}\label{eq:scale-score}
\frac{S_w(\hat{\mu}_w)}{W}=m_{2,d}(\hat{\beta}_w).
\end{equation}
In other words, the inverse-scale parameter is estimated by matching the weighted empirical mean squared radius to the model-implied radial second moment. Since \(m_{2,d}(\beta)\) is strictly decreasing by Proposition~\ref{prop:Ad-derivatives}, the profile equation \eqref{eq:scale-score} has at most one solution whenever it is well defined.

We emphasize that the weighted formulation in \eqref{eq:weighted-loglik} is not a technical side issue. It is the exact optimization problem that appears inside the M-step of the finite-mixture model developed later. Consequently, once the weighted barycenter problem and the one-dimensional scale profile are understood, the main statistical and computational ingredients of the mixture model are already in place.

%%%%%%%%%%%%%%%%%%%%%%%%%%%%%%%%%%%%%%%%%%%%%%%%%%%%%%%%%%%%%%%%%%%%%%%%%%%%%%%%%
%%%%%%%%%%%%%%%%%%%%%%%%%%%%%%%%%%%%%%%%%%%%%%%%%%%%%%%%%%%%%%%%%%%%%%%%%%%%%%%%%
\section{Finite Riemannian Gaussian mixtures and computation}\label{sec:computation}

Having developed the weighted single-component estimation problem, we now turn to finite mixture modeling. The key point is that the weighted formulation is not ancillary to the mixture problem. Rather, it is the mixture problem at the level of one component and one M-step. Once posterior responsibilities are introduced, every component-wise location update becomes a weighted barycenter problem, and every component-wise scale update becomes a one-dimensional weighted profile optimization. For this reason, the computational and theoretical structure of the finite-mixture model is already implicit in the weighted likelihood of Section~\ref{sec:model}.

The present section develops the finite-mixture formulation and the algorithms used to fit it. We begin with the complete-data representation underlying EM, then explain why a constrained parameter space is needed, and finally describe the exact EM and generalized EM schemes used in computation. Since the dominant numerical cost lies in the repeated evaluation of weighted Fr\'echet means, we also present an explicit hyperbolic majorization--minimization routine for that subproblem and explain how it interfaces with both EM and GEM.

\subsection{Finite mixtures and the complete-data representation}\label{subsec:mixture-definition}

Let \(K\ge 2\) be the pre-specified number of components. A finite mixture of Riemannian Gaussian distributions on \(\mathbb{H}^d\) has density
\begin{equation}\label{eq:mixture-density}
f(x;\theta)
=
\sum_{k=1}^K \pi_k\, p(x\mid \mu_k,\beta_k),
\qquad
\theta=\{(\pi_k,\mu_k,\beta_k):1\le k\le K\},
\end{equation}
where \(\pi_k\ge 0\) and \(\sum_{k=1}^K \pi_k=1\). Here \(\pi_k\) is the mixing weight, \(\mu_k\in\mathbb{H}^d\) is the location parameter, and \(\beta_k>0\) is the inverse-scale parameter of the $k$-th component. Given observations \(x_1,\ldots,x_n\in\mathbb{H}^d\), the observed-data log-likelihood is
\begin{equation}\label{eq:observed-loglik}
\ell(\theta)
=
\sum_{i=1}^n
\log\Bigg[
\sum_{k=1}^K
\pi_k
\exp\!\Big\{
-A_d(\beta_k)-\beta_k d^2(x_i,\mu_k)
\Big\}
\Bigg].
\end{equation}

As in ordinary finite-mixture theory, it is convenient to introduce latent membership variables \(z_{ik}\in\{0,1\}\), where \(z_{ik}=1\) indicates that observation \(x_i\) belongs to component \(k\) and \(\sum_{k=1}^K z_{ik}=1\). The corresponding complete-data log-likelihood is
\begin{equation}\label{eq:complete-data-loglik}
\ell_c(\theta)
=
\sum_{i=1}^n\sum_{k=1}^K
z_{ik}
\Big[
\log\pi_k-A_d(\beta_k)-\beta_k d^2(x_i,\mu_k)
\Big].
\end{equation}
This is the standard missing-data formulation underlying the EM algorithm \citep{dempster_1977_MaximumLikelihoodIncomplete}. In the present model, once the current posterior probabilities \(r_{ik}\) are computed, they enter the M-step exactly as weights. Thus the weighted estimation problem studied in Section~\ref{sec:model} reappears verbatim inside every outer EM iteration.

\subsection{Why a constrained mixture model is needed}\label{subsec:constrained-mixture}

Before turning to computation, it is important to recognize that the unrestricted finite-mixture likelihood is not well posed. This is not a defect of the Riemannian Gaussian model itself, but rather the hyperbolic analogue of the familiar singularity of Euclidean Gaussian mixtures. That is, if one component is allowed to become arbitrarily concentrated around a single data point, the observed likelihood can diverge.

\begin{proposition}[Unrestricted finite-mixture singularity]\label{prop:mixture-singularity}
Suppose \(K\ge 2\). If the component parameters \(\mu_k\in\mathbb{H}^d\), \(\beta_k>0\), and \(\pi_k\ge 0\) are unconstrained except for \(\sum_{k=1}^K \pi_k=1\), then the observed log-likelihood \eqref{eq:observed-loglik} is unbounded above.
\end{proposition}

\begin{proof}
Fix one observation, say \(x_1\). Choose a component with parameters
\[
\pi_1=\varepsilon\in(0,1),
\qquad
\mu_1=x_1,
\qquad
\beta_1\to\infty,
\]
and choose the remaining mixture mass so that the other observations continue to receive positive density, for instance via any fixed background component. At \(x_1\), the first component contributes
\[
\pi_1 p(x_1\mid x_1,\beta_1)
=
\varepsilon Z_d(\beta_1)^{-1}.
\]
By Proposition~\ref{prop:Zd-asymptotic},
\[
Z_d(\beta_1)^{-1}\sim \pi^{-d/2}\beta_1^{d/2}\to\infty.
\]
Hence the contribution of the first observation to \(\ell(\theta)\) diverges to \(+\infty\), while the remaining terms stay finite. Therefore the observed log-likelihood is unbounded above.
\end{proof}

Proposition~\ref{prop:mixture-singularity} is the hyperbolic counterpart of the usual Gaussian-mixture collapse in Euclidean space. The phenomenon is model based rather than coordinate based. It arises because a component density can become arbitrarily concentrated near a single observation. Consequently, any practical finite-mixture methodology must either constrain, penalize, or regularize the scale parameters.

To obtain a well-posed estimation problem, we restrict attention to a compact parameter space. Let
\begin{equation}\label{eq:Cn-def}
C_n
:=
\overline{\operatorname{conv}_g\{x_1,\ldots,x_n\}}    
\end{equation}
denote the closed geodesic convex hull of the sample and 
$
0<\underline{\beta}<\overline{\beta}<\infty
$
be fixed scale bounds. We define
\begin{equation}\label{eq:theta-c-def}
\Theta_c
=
\Big\{
\theta=(\pi,\mu,\beta):
\ \pi\in\Delta_{K-1},\ 
\mu_k\in C_n,\ 
\beta_k\in[\underline{\beta},\overline{\beta}],
\ 1\le k\le K
\Big\},
\end{equation}
where \(\Delta_{K-1}=\{\pi\in[0,1]^K:\sum_{k=1}^K \pi_k=1\}\) is the closed \((K-1)\)-simplex.

This restriction is natural for both statistical and geometric reasons. First, exact location updates are weighted \Frechet means and therefore lie in the geodesic convex hull of the positively weighted sample points \citep{afsari_2011_Riemannian$L_p$Center}. Second, constraining \(\beta_k\) to a compact interval prevents scale collapse. Third, the resulting parameter space is compact, which will later be used to establish existence of the constrained mixture estimator and continuity of the EM map.

\subsection{A common computational template for exact EM and GEM}\label{subsec:common-em-gem}

The constrained mixture model admits a particularly transparent block structure. At outer iteration \(t\), given the current parameter \(\theta^{(t)}\), the posterior responsibility of component \(k\) for observation \(x_i\) is
\begin{equation}\label{eq:responsibilities}
r_{ik}^{(t)}
=
\frac{
\pi_k^{(t)}
\exp\!\big[
-A_d(\beta_k^{(t)})
-\beta_k^{(t)} d^2(x_i,\mu_k^{(t)})
\big]
}{
\sum_{j=1}^K
\pi_j^{(t)}
\exp\!\big[
-A_d(\beta_j^{(t)})
-\beta_j^{(t)} d^2(x_i,\mu_j^{(t)})
\big]
}.
\end{equation}
For numerical stability, it is preferable to compute these quantities in logarithmic form. Define
\begin{equation}\label{eq:eta-ik}
\eta_{ik}^{(t)}
=
\log \pi_k^{(t)}
-
A_d(\beta_k^{(t)})
-
\beta_k^{(t)}d^2(x_i,\mu_k^{(t)}),
\end{equation}
and then evaluate
\begin{equation}\label{eq:log-responsibility}
\log r_{ik}^{(t)}
=
\eta_{ik}^{(t)}-\operatorname{LSE}_{1\le j\le K}\eta_{ij}^{(t)},
\end{equation}
where \(\operatorname{LSE}\) denotes the log-sum-exp operator.

Once the responsibilities are available, the effective component weights and weighted within-component variance functionals are given as 
\begin{equation}
W_k^{(t)}=\sum_{i=1}^n r_{ik}^{(t)}\quad\text{and}\quad S_k^{(t)}(\mu)
=
\sum_{i=1}^n r_{ik}^{(t)}\,d^2(x_i,\mu),
\end{equation}
respectively. Then, the EM surrogate criterion takes the block-separable form
\begin{equation}\label{eq:Q-block-structure}
\mathcal{Q}(\theta\mid\theta^{(t)})
=
\sum_{k=1}^K
\Big[
W_k^{(t)}\log\pi_k
-
W_k^{(t)}A_d(\beta_k)
-
\beta_k S_k^{(t)}(\mu_k)
\Big].
\end{equation}
This identity makes the structure of the M-step completely explicit: the simplex block, the \(K\) location blocks, and the \(K\) one-dimensional scale blocks decouple.

At this point, the distinction between exact EM and generalized EM (GEM) becomes purely computational. In this paper we use the following terminology.
\begin{enumerate}[label=(\roman*)]
    \item \textbf{Exact EM} means that for each component \(k\), the location block is solved exactly:
    \[
    \mu_k^{(t+1)}=\arg\min_{\mu\in C_n} S_k^{(t)}(\mu),
    \]
    while the mixing-weight and scale blocks are also solved exactly.
    \item \textbf{GEM} means that the same E-step is used, the mixing-weight and scale blocks are still solved exactly, but the location block is replaced by any update \(\widetilde{\mu}_k^{(t+1)}\) satisfying
    \[
    S_k^{(t)}\bigl(\widetilde{\mu}_k^{(t+1)}\bigr)
    \le
    S_k^{(t)}\bigl(\mu_k^{(t)}\bigr).
    \]
    Thus the location block need only improve the surrogate criterion, not maximize it exactly.
\end{enumerate}

This formulation is more precise than the generic statement saying that  GEM uses approximate M-steps. In the present problem, the approximation is isolated to one block only: the weighted barycenter computation. The simplex update and the scale update remain exact in both algorithms. This observation will be important later because all monotonicity arguments reduce to the simple fact that decreasing \(S_k^{(t)}\) increases the $k$-th  component's contribution to \(\mathcal{Q}(\theta\mid\theta^{(t)})\).

\subsection{Exact EM algorithm}\label{subsec:exact-em}

With the preceding block decomposition in place, the exact EM algorithm becomes straightforward. At iteration \(t\), one maximizes \(\mathcal{Q}(\theta\mid\theta^{(t)})\) over the constrained parameter space \(\Theta_c\). Since the simplex block and the scale blocks are explicit, exactness reduces to the requirement that each weighted Fr\'echet-mean subproblem be solved to its unique minimizer.

The exact M-step is therefore
\begin{align}
\pi_k^{(t+1)}
&=
\frac{W_k^{(t)}}{n},
\label{eq:pi-update}
\\
\mu_k^{(t+1)}
&=
\arg\min_{\mu\in C_n} S_k^{(t)}(\mu),
\label{eq:mu-update}
\\
\beta_k^{(t+1)}
&=
\arg\min_{\beta\in[\underline{\beta},\overline{\beta}]}
\Big\{
\beta S_k^{(t)}(\mu_k^{(t+1)})
+
W_k^{(t)}A_d(\beta)
\Big\}.
\label{eq:beta-update}
\end{align}
Thus exact EM is entirely explicit once a reliable routine for the weighted barycenter problem is available.

\begin{algorithm}[t]
\caption{Exact EM for constrained Riemannian Gaussian mixtures}
\label{alg:exact-em}
\begin{algorithmic}[1]
\REQUIRE Data \(x_1,\ldots,x_n\in\mathbb{H}^d\); number of components \(K\); scale bounds \([\underline{\beta},\overline{\beta}]\); initial parameter \(\theta^{(0)}\in\Theta_c\) with \(\pi_k^{(0)}>0\); tolerance \(\varepsilon_{\mathrm{out}}>0\)
\ENSURE Final iterate \(\theta^{(t)}\)

\STATE Optionally precompute \(A_d\), \(A_d'\), and \(A_d''\)
\STATE Set \(t\gets 0\)

\WHILE{the stopping criterion is not met}
    \STATE Compute the responsibilities \(r_{ik}^{(t)}\)
    \STATE Compute the effective weights \(W_k^{(t)}\)
    \STATE Update \(\pi_k^{(t+1)}=W_k^{(t)}/n\) for all \(k\)

    \FOR{$k=1,\ldots,K$}
        \STATE Update \(\mu_k^{(t+1)}\) by solving the weighted barycenter problem in \eqref{eq:mu-update}
        \STATE Update \(\beta_k^{(t+1)}\) by solving the one-dimensional convex problem in \eqref{eq:beta-update}
    \ENDFOR

    \STATE Evaluate \(\ell(\theta^{(t+1)})\)
    \STATE Stop if the criterion in the text is satisfied
    \STATE Set \(t\gets t+1\)
\ENDWHILE
\end{algorithmic}
\end{algorithm}

The main conceptual point is that Algorithm~\ref{alg:exact-em} is exact only because the location subproblems are solved completely. Every other block in the M-step already has a closed or low-dimensional form. This one-block distinction will later make the comparison with GEM especially transparent: to pass from exact EM to GEM, one changes only the location update and leaves every other part of the algorithm untouched.

\subsection{Weighted Fr\'echet-mean computation via hyperbolic MM}\label{subsec:mm-barycenter}

The dominant numerical cost in the EM algorithm is the repeated solution of the location update \eqref{eq:mu-update}. Since the weighted \Frechet objective is strongly geodesically convex, generic Riemannian gradient descent would already be valid \citep{boumal_2023_IntroductionOptimizationSmooth}. However, inside an EM loop one prefers an update that is explicit, hyperparameter free, inexpensive per iteration, and naturally adapted to the hyperboloid model. We therefore use a hyperbolic majorization--minimization (MM) routine, following the same computational logic advocated by \cite{lou_2020_DifferentiatingFrechetMean}.

Recall the weighted barycenter objective
\begin{equation}\label{eq:barycenter-objective}
f(\mu)
=
\sum_{i=1}^n w_i\,\operatorname{arcosh}^2\bigl(-\langle x_i,\mu\rangle_L\bigr),
\qquad \mu\in\mathbb{H}^d,
\end{equation}
where \(w_i\ge 0\) and \(W=\sum_{i=1}^n w_i>0\). Define also
\begin{equation}\label{eq:mm-phi-def}
\phi(a)
=
\frac{2\operatorname{arcosh}(a)}{\sqrt{a^2-1}},
\qquad a>1,
\end{equation}
with the continuous extension \(\phi(1)=2\). The MM construction rests on the concavity of \(a\mapsto \operatorname{arcosh}^2(a)\).

\begin{lemma}[Concavity of \(a\mapsto \operatorname{arcosh}^2(a)\)]\label{lem:mm-concavity}
The function
\[
g(a):=\operatorname{arcosh}^2(a),\qquad a\ge 1,
\]
is continuously differentiable on \([1,\infty)\), twice continuously differentiable on \((1,\infty)\), and concave on \([1,\infty)\). Moreover,
\begin{equation}\label{eq:mm-gprime}
g'(a)=\frac{2\operatorname{arcosh}(a)}{\sqrt{a^2-1}}=\phi(a),
\qquad a>1.
\end{equation}
\end{lemma}

\begin{proof}
For \(a>1\), differentiation of
\[
\operatorname{arcosh}(a)=\log\!\bigl(a+\sqrt{a^2-1}\bigr)
\]
gives
\[
\frac{d}{da}\operatorname{arcosh}(a)=\frac{1}{\sqrt{a^2-1}},
\]
and therefore
\[
g'(a)=2\operatorname{arcosh}(a)\frac{1}{\sqrt{a^2-1}}=\phi(a).
\]
Differentiating again yields
\[
g''(a)
=
\frac{2\sqrt{a^2-1}-2a\operatorname{arcosh}(a)}{(a^2-1)^{3/2}}.
\]
To determine the sign, write \(a=\cosh t\) with \(t\ge 0\). Then \(\sqrt{a^2-1}=\sinh t\) and \(\operatorname{arcosh}(a)=t\), so
\[
\sqrt{a^2-1}-a\operatorname{arcosh}(a)
=
\sinh t-t\cosh t
=
-\cosh t\,(t-\tanh t).
\]
Now \(h(t)=t-\tanh t\) satisfies \(h(0)=0\) and
\[
h'(t)=1-\operatorname{sech}^2(t)=\tanh^2(t)\ge 0,
\]
hence \(h(t)\ge 0\) for all \(t\ge 0\). Therefore \(\sqrt{a^2-1}-a\operatorname{arcosh}(a)\le 0\), which implies \(g''(a)\le 0\). Thus \(g\) is concave on \([1,\infty)\).
\end{proof}

The next step is to show that the coefficient vector arising from the tangent-line majorization remains future timelike.

\begin{lemma}[The MM coefficient vector is future timelike]\label{lem:mm-timelike}
Fix \(\mu\in\mathbb{H}^d\), and define
\begin{equation}\label{eq:mm-ci-def}
\alpha_i(\mu):=-\langle x_i,\mu\rangle_L\ge 1,
\qquad
c_i(\mu):=w_i\phi\bigl(\alpha_i(\mu)\bigr),
\qquad
\nu(\mu):=\sum_{i=1}^n c_i(\mu)x_i.
\end{equation}
Then \(\nu(\mu)\) is future timelike, that is,
\[
\langle \nu(\mu),\nu(\mu)\rangle_L<0,
\qquad
\nu_{d+1}(\mu)>0.
\]
\end{lemma}

\begin{proof}
Because \(W>0\), at least one weight is strictly positive. Since \(\phi(a)>0\) for all \(a\ge 1\), at least one coefficient \(c_i(\mu)\) is strictly positive and all are nonnegative. Every point \(x_i\in\mathbb{H}^d\) has positive last coordinate, so
\[
\nu_{d+1}(\mu)=\sum_{i=1}^n c_i(\mu)(x_i)_{d+1}>0.
\]
For the Lorentzian norm, observe that
\[
-\langle x_i,x_j\rangle_L=\cosh d(x_i,x_j)\ge 1,
\]
hence
\[
\langle \nu(\mu),\nu(\mu)\rangle_L
=
\sum_{i=1}^n\sum_{j=1}^n c_i(\mu)c_j(\mu)\langle x_i,x_j\rangle_L
\le
-\sum_{i=1}^n\sum_{j=1}^n c_i(\mu)c_j(\mu)
=
-\left(\sum_{i=1}^n c_i(\mu)\right)^2<0.
\]
Therefore \(\nu(\mu)\) is future timelike.
\end{proof}

Together, these ingredients lead to a particularly simple global majorizer.

\begin{proposition}[A global linear majorizer]\label{prop:mm-majorizer}
Fix a current iterate \(\mu^{(s)}\in\mathbb{H}^d\), and define
\[
\alpha_i^{(s)}:=-\langle x_i,\mu^{(s)}\rangle_L,
\qquad
c_i^{(s)}:=w_i\phi\bigl(\alpha_i^{(s)}\bigr),
\qquad
\nu^{(s+1)}:=\sum_{i=1}^n c_i^{(s)}x_i.
\]
Let
\begin{equation}\label{eq:mm-majorizer}
M_s(\mu)
:=
\sum_{i=1}^n
w_i
\Big[
g\bigl(\alpha_i^{(s)}\bigr)
+
g'\bigl(\alpha_i^{(s)}\bigr)
\bigl(-\langle x_i,\mu\rangle_L-\alpha_i^{(s)}\bigr)
\Big].
\end{equation}
Then:
\begin{enumerate}[label=(\roman*)]
    \item \(f(\mu)\le M_s(\mu)\) for every \(\mu\in\mathbb{H}^d\), with equality at \(\mu=\mu^{(s)}\).
    \item Up to an additive constant independent of \(\mu\), the majorizer is linear in \(\mu\):
    \begin{equation}\label{eq:mm-majorizer-linear}
    M_s(\mu)=\mathrm{const}-\langle \nu^{(s+1)},\mu\rangle_L.
    \end{equation}
    \item The unique minimizer of \(M_s\) over \(\mathbb{H}^d\) is
    \begin{equation}\label{eq:mm-update-explicit}
    \mu^{(s+1)}
    =
    \frac{\nu^{(s+1)}}{\sqrt{-\langle \nu^{(s+1)},\nu^{(s+1)}\rangle_L}}.
    \end{equation}
\end{enumerate}
\end{proposition}

\begin{proof}
By concavity of \(g\), every tangent line majorizes \(g\):
\[
g(a)\le g(a_0)+g'(a_0)(a-a_0)
\qquad\text{for all }a,a_0\ge 1.
\]
Applying this inequality with \(a=-\langle x_i,\mu\rangle_L\) and \(a_0=\alpha_i^{(s)}\), then summing over \(i\), proves part (i).

For part (ii), note that all dependence on \(\mu\) is contained in
\[
-\sum_{i=1}^n w_i g'\bigl(\alpha_i^{(s)}\bigr)\langle x_i,\mu\rangle_L
=
-\sum_{i=1}^n c_i^{(s)}\langle x_i,\mu\rangle_L
=
-\langle \nu^{(s+1)},\mu\rangle_L,
\]
which gives \eqref{eq:mm-majorizer-linear}.

For part (iii), Lemma~\ref{lem:mm-timelike} shows that \(\nu^{(s+1)}\) is future timelike. The reverse Lorentzian Cauchy--Schwarz inequality implies that for every future timelike \(u\) and every \(\mu\in\mathbb{H}^d\),
\[
-\langle u,\mu\rangle_L\ge \sqrt{-\langle u,u\rangle_L},
\]
with equality if and only if \(\mu=u/\sqrt{-\langle u,u\rangle_L}\). Applying this with \(u=\nu^{(s+1)}\) proves \eqref{eq:mm-update-explicit}.
\end{proof}

The resulting MM routine is completely explicit, which is summarized in Algorithm~\ref{alg:mm-barycenter}.
\begin{algorithm}[t]
\caption{MM update for the weighted hyperbolic Fr\'echet mean}
\label{alg:mm-barycenter}
\begin{algorithmic}[1]
\REQUIRE Weighted data \((x_i,w_i)_{i=1}^n\subset \mathbb{H}^d\); initial iterate \(\mu^{(0)}\in\mathbb{H}^d\)
\ENSURE Approximate weighted Fr\'echet mean \(\mu^{(s)}\)

\STATE Set \(s\gets 0\)

\WHILE{not converged}
    \STATE Compute the coefficients \(\alpha_i^{(s)}\), \(c_i^{(s)}\), and the ambient vector \(\nu^{(s+1)}\) as in Proposition~\ref{prop:mm-majorizer}
    \STATE Update \(\mu^{(s+1)}\) using \eqref{eq:mm-update-explicit}
    \STATE Set \(s\gets s+1\)
\ENDWHILE

\end{algorithmic}
\end{algorithm}

The next theorem shows that the MM map is well behaved as an optimization routine for the weighted barycenter problem.

\begin{theorem}[Monotonic descent and convergence of the MM iterates]\label{thm:mm-convergence}
Let \(\{\mu^{(s)}\}_{s\ge 0}\) be the sequence generated by Algorithm~\ref{alg:mm-barycenter}. Then:
\begin{enumerate}[label=(\roman*)]
    \item The objective values are monotone:
    \begin{equation}\label{eq:mm-descent}
    f\bigl(\mu^{(s+1)}\bigr)\le f\bigl(\mu^{(s)}\bigr)
    \qquad\text{for all }s\ge 0,
    \end{equation}
    with strict inequality whenever \(\mu^{(s+1)}\neq \mu^{(s)}\).
    \item The sequence \(\{f(\mu^{(s)})\}\) converges, and the iterates remain in a compact sublevel set of \(\mathbb{H}^d\).
    \item Every accumulation point of \(\{\mu^{(s)}\}\) is a fixed point of the MM map
    \begin{equation}\label{eq:mm-map-def}
    \mathcal{T}_{\mathrm{MM}}(\mu)
    :=
    \frac{\nu(\mu)}{\sqrt{-\langle \nu(\mu),\nu(\mu)\rangle_L}}.
    \end{equation}
    \item A point \(\mu\in\mathbb{H}^d\) is a fixed point of \(\mathcal{T}_{\mathrm{MM}}\) if and only if it satisfies the weighted score equation
    \begin{equation}\label{eq:mm-fixedpoint-score}
    \sum_{i=1}^n w_i\,\operatorname{Log}_\mu(x_i)=0.
    \end{equation}
\end{enumerate}
Consequently, once uniqueness of the weighted Fr\'echet mean is invoked, the full MM sequence converges to that unique barycenter.
\end{theorem}

\begin{proof}
By part (i) of Proposition~\ref{prop:mm-majorizer},
\[
f\bigl(\mu^{(s+1)}\bigr)\le M_s\bigl(\mu^{(s+1)}\bigr),
\]
and by part (iii), \(\mu^{(s+1)}\) minimizes \(M_s\). Therefore
\[
M_s\bigl(\mu^{(s+1)}\bigr)\le M_s\bigl(\mu^{(s)}\bigr)=f\bigl(\mu^{(s)}\bigr),
\]
which proves \eqref{eq:mm-descent}. If \(\mu^{(s+1)}\neq \mu^{(s)}\), then \(\mu^{(s+1)}\) is the unique minimizer of \(M_s\), so the inequality is strict.

Because \(f\ge 0\), the monotone sequence \(\{f(\mu^{(s)})\}\) converges to a finite limit. Moreover, if \(j\) is any index with \(w_j>0\), then
\[
f(\mu)\ge w_j d^2(x_j,\mu).
\]
Hence every sublevel set \(\{\mu:f(\mu)\le c\}\) is bounded. Since \(\mathbb{H}^d\) is proper, bounded closed sets are compact, which proves part (ii).

For part (iii), note first that \(\phi\) is continuous on \([1,\infty)\), so \(\nu(\mu)\) and \(\mathcal{T}_{\mathrm{MM}}(\mu)\) are continuous on \(\mathbb{H}^d\). Let \(\bar{\mu}\) be an accumulation point, and choose a subsequence \(\mu^{(s_m)}\to\bar{\mu}\). Suppose for contradiction that \(\mathcal{T}_{\mathrm{MM}}(\bar{\mu})\neq \bar{\mu}\). Then the strict descent property implies
\[
\Delta(\mu):=f(\mu)-f\bigl(\mathcal{T}_{\mathrm{MM}}(\mu)\bigr)>0
\qquad\text{at }\mu=\bar{\mu}.
\]
By continuity of \(\Delta\), there exist \(\varepsilon>0\) and \(m_0\) such that \(\Delta(\mu^{(s_m)})\ge \varepsilon\) for all \(m\ge m_0\). But then
\[
f\bigl(\mu^{(s_m)}\bigr)-f\bigl(\mu^{(s_m+1)}\bigr)
\ge
\Delta\bigl(\mu^{(s_m)}\bigr)
\ge
\varepsilon,
\]
contradicting convergence of \(f(\mu^{(s)})\). Hence every accumulation point is a fixed point of \(\mathcal{T}_{\mathrm{MM}}\).

For part (iv), suppose first that \(\mu\) is a fixed point of \(\mathcal{T}_{\mathrm{MM}}\). Then \(\nu(\mu)=\lambda\mu\) for some scalar \(\lambda>0\). Taking Lorentzian inner products with \(\mu\) gives
\[
\lambda=-\langle \nu(\mu),\mu\rangle_L=\sum_{i=1}^n c_i(\mu)\alpha_i(\mu),
\]
so
\[
0
=
\nu(\mu)-\lambda\mu
=
\sum_{i=1}^n c_i(\mu)\bigl(x_i-\alpha_i(\mu)\mu\bigr).
\]
Dividing by two and recalling the definitions of \(c_i(\mu)\) and \(\operatorname{Log}_\mu(x_i)\) yields \eqref{eq:mm-fixedpoint-score}. Conversely, if \eqref{eq:mm-fixedpoint-score} holds, then the same identity shows
\[
\nu(\mu)
=
\left(
\sum_{i=1}^n c_i(\mu)\alpha_i(\mu)
\right)\mu,
\]
with positive scalar coefficient, so normalization returns \(\mu\). Hence \(\mu\) is a fixed point of \(\mathcal{T}_{\mathrm{MM}}\).

Finally, since $\mathbb{H}^d$ is a Hadamard manifold, the weighted \Frechet mean exists and is unique \citep{afsari_2011_Riemannian$L_p$Center}. Hence the weighted score equation \eqref{eq:mm-fixedpoint-score} admits a unique solution. Since every accumulation point of the MM sequence satisfies this equation, all accumulation points coincide with that unique solution, and therefore the full sequence converges to the weighted \Frechet mean.
\end{proof}

A few implementation remarks are in order. Warm starts are effective in practice: initializing each component mean at iteration \(t+1\) by \(\mu_k^{(t)}\) typically reduces the number of inner MM iterations. Numerical stabilization is required when \(\alpha\downarrow 1\), since \(\operatorname{arcosh}(\alpha)/\sqrt{\alpha^2-1}\to 1\). In practice, one may clip \(\alpha\) to \(1+\varepsilon\) or use a short series expansion near \(\alpha=1\). Finally, each MM iteration is inexpensive, requiring only Lorentzian inner products and a single weighted sum, yielding an overall cost of order \(O(nd)\).

\subsection{One-dimensional scale updates and precomputation}\label{subsec:scale-computation}

Once the location parameter is fixed, the scale update is a scalar optimization problem. For known \(S\ge 0\) and \(W>0\), the generic subproblem is
\begin{equation}\label{eq:beta-subproblem}
\min_{\beta\in[\underline{\beta},\overline{\beta}]}
\Big\{
\beta S+W A_d(\beta)
\Big\}.
\end{equation}
By Proposition~\ref{prop:Ad-derivatives}, this objective is strictly convex. The corresponding unconstrained first-order equation is
\begin{equation*}
m_{2,d}(\beta)=S/W.
\end{equation*}
Thus the scale update can be interpreted as a radial moment-matching problem.

To solve the problem, we employ a modified Newton method, which can be derived from the first-order optimality condition for \eqref{eq:beta-subproblem}. Differentiating the objective with respect to \(\beta\) yields
\[
g'(\beta)=S+W A_d'(\beta),
\]
so that the unconstrained optimum satisfies the moment-matching equation \(A_d'(\beta)=-S/W\). Applying Newton's method to the root-finding problem \(g'(\beta)=0\) gives the iteration
\[
\beta \;\leftarrow\; \beta-\frac{g'(\beta)}{g''(\beta)}
=
\beta-\frac{A_d'(\beta)+S/W}{A_d''(\beta)},
\]
where we used \(g''(\beta)=W A_d''(\beta)\). To respect the constraint \(\beta\in[\underline{\beta},\overline{\beta}]\), this update is followed by projection onto the interval, yielding
\begin{equation}\label{eq:newton-beta}
\beta^{\mathrm{new}}
=
\Pi_{[\underline{\beta},\overline{\beta}]}
\left[
\beta-\frac{A_d'(\beta)+S/W}{A_d''(\beta)}
\right].
\end{equation}
Since the objective is one dimensional and strictly convex by Proposition~\ref{prop:Ad-derivatives}, the Newton iteration is well behaved and converges rapidly in practice. Consequently, the scale update is computationally negligible compared with the barycenter updates in the M-step.

In repeated mixture fitting, the most effective implementation strategy is to precompute the scalar functions \(A_d\), \(A_d'\), and \(A_d''\) on a grid of \(\beta\)-values once per ambient dimension \(d\), and then use interpolation together with one or two Newton refinements. This does not change the statistical problem, but it can substantially reduce the per-iteration computational cost. Since the dimension is fixed across all components and all outer iterations, the same precomputed table can be reused globally.

\subsection{A generalized EM variant with truncated inner barycenter solves}\label{subsec:gem}

Generalized EM replaces exact maximization of the surrogate criterion by updates that merely increase it. In the present model, there is no reason to approximate the entire M-step. The mixing-weight block is closed form, the scale block is one dimensional and strictly convex, and only the weighted Fr\'echet-mean block is computationally demanding. Our GEM formulation is therefore deliberately conservative: it keeps the E-step, the effective weights, the simplex update, and the scale update exactly the same as in Algorithm~\ref{alg:exact-em}, and relaxes only the location update.

More precisely, after the E-step at iteration \(t\), fix a component \(k\) and consider the weighted variance
\[
S_k^{(t)}(\mu)=\sum_{i=1}^n r_{ik}^{(t)}d^2(x_i,\mu).
\]
Exact EM computes the unique minimizer of this function over \(C_n\). GEM instead starts from the previous mean \(\mu_k^{(t)}\) and applies only a finite number of descent iterations. In this paper, the descent routine is the MM barycenter map from Algorithm~\ref{alg:mm-barycenter}. Thus, if \(L_k^{(t)}\ge 1\) denotes the chosen inner-step budget for component \(k\) at outer iteration \(t\), we define
\[
\mu_{k,0}^{(t+1)}:=\mu_k^{(t)},
\qquad
\mu_{k,\ell+1}^{(t+1)}
:=
\mathcal{T}_{\mathrm{MM},k}^{(t)}\bigl(\mu_{k,\ell}^{(t+1)}\bigr),
\qquad
\ell=0,\ldots,L_k^{(t)}-1,
\]
where \(\mathcal{T}_{\mathrm{MM},k}^{(t)}\) denotes the MM map associated with the weighted objective \(S_k^{(t)}\). The GEM location update is then
\begin{equation}\label{eq:gem-location-update}
\mu_k^{(t+1)}:=\mu_{k,L_k^{(t)}}^{(t+1)}.
\end{equation}
Because every MM step decreases the corresponding weighted variance, the resulting update automatically satisfies the descent condition required later in the GEM monotonicity theorem.

This formulation makes the relationship between exact EM and GEM particularly transparent.
\begin{itemize}[leftmargin=2.1em]
    \item If each \(L_k^{(t)}\) is large enough that the inner barycenter solver reaches numerical convergence, then GEM reduces to exact EM.
    \item If each \(L_k^{(t)}=1\), one obtains the cheapest admissible GEM update: a single MM step per component per outer iteration.
    \item Intermediate choices of \(L_k^{(t)}\) interpolate between these two extremes and allow a direct trade-off between per-iteration cost and the aggressiveness of the M-step.
\end{itemize}
Thus GEM is not a different statistical model, but rather a computationally budgeted version of the same EM framework.

\begin{algorithm}[t]
\caption{Generalized EM with truncated MM barycenter updates}
\label{alg:gem}
\begin{algorithmic}[1]
\REQUIRE Same inputs as Algorithm~\ref{alg:exact-em}, together with an inner-step schedule \(L_k^{(t)}\ge 1\) for each component and outer iteration
\ENSURE Final iterate \(\theta^{(t)}\)

\STATE Optionally precompute \(A_d\), \(A_d'\), and \(A_d''\)
\STATE Set \(t\gets 0\)

\WHILE{the stopping criterion is not met}
    \STATE Compute \(\eta_{ik}^{(t)}\) and the responsibilities \(r_{ik}^{(t)}\) using \eqref{eq:eta-ik} and \eqref{eq:log-responsibility}
    \STATE Compute the effective weights \(W_k^{(t)}\) for \(k=1,\ldots,K\)
    \STATE Update \(\pi_k^{(t+1)}=W_k^{(t)}/n\) for all \(k\)

    \FOR{$k=1,\ldots,K$}
        \STATE Initialize the inner loop at \(\mu_{k,0}^{(t+1)}=\mu_k^{(t)}\)
        \STATE Apply \(L_k^{(t)}\) MM iterations of Algorithm~\ref{alg:mm-barycenter} to the weighted objective \(S_k^{(t)}\)
        \STATE Set \(\mu_k^{(t+1)}\) according to \eqref{eq:gem-location-update}
        \STATE Update \(\beta_k^{(t+1)}\) by solving the one-dimensional convex problem in \eqref{eq:beta-update}
    \ENDFOR

    \STATE Evaluate \(\ell(\theta^{(t+1)})\)
    \STATE Stop if the criterion in the text is satisfied
    \STATE Set \(t\gets t+1\)
\ENDWHILE

\end{algorithmic}
\end{algorithm}

The structural similarity between Algorithms~\ref{alg:exact-em} and \ref{alg:gem} is intentional. The two procedures share the same state variables, the same E-step, the same mixing-weight update, and the same scale update. They differ only in the extent to which the location block is optimized before returning to the E-step. This is precisely why the monotonicity proof for GEM is so simple later on: one only needs to verify that the truncated inner loop does not increase \(S_k^{(t)}\).

From a practical standpoint, exact EM is preferable when one wants the cleanest outer-iteration theory and is willing to solve each barycenter problem essentially to completion. GEM is preferable when the number of components is large, the sample is large, or one wishes to cap the inner computational effort in a predictable way. The two can also be hybridized: one may use GEM with small \(L_k^{(t)}\) during early exploratory iterations and switch to exact EM for final refinement once the responsibilities stabilize.

%%%%%%%%%%%%%%%%%%%%%%%%%%%%%%%%%%%%%%%%%%%%%%%%%%%%%%%%%%%%%%%%%%%%%%%%%%%%%%%%%
%%%%%%%%%%%%%%%%%%%%%%%%%%%%%%%%%%%%%%%%%%%%%%%%%%%%%%%%%%%%%%%%%%%%%%%%%%%%%%%%%
\section{Theory}\label{sec:theory}

This section collects the main theoretical properties supporting the methodology developed in Sections~\ref{sec:model} and \ref{sec:computation}. We begin with the weighted single-component problem and show that the Riemannian Gaussian maximum likelihood estimator is well posed. We then turn to the finite-mixture model, where we establish existence of the constrained estimator. Finally, we study the optimization procedures, proving monotonicity of the exact EM algorithm, continuity of its update map under a mild interiority condition, and likelihood ascent for the generalized EM scheme based on truncated inner barycenter updates.

\subsection{Existence and uniqueness of the weighted single-component MLE}\label{subsec:single-mle-theory}

We first record the convexity property that underlies both barycenter uniqueness and likelihood-based estimation. Since \(\mathbb{H}^d\) is a Hadamard manifold, the squared distance is strongly geodesically convex along geodesics; see \eqref{eq:bg_strong_convexity}. Summing this inequality with weights immediately yields the corresponding property for the weighted Fr\'echet functional.

\begin{theorem}[Strong geodesic convexity of the weighted variance]\label{thm:Sw-strong-convexity}
Let \(S_w\) be defined by \eqref{eq:weighted-variance}. Then, for every geodesic \(\gamma:[0,1]\to\mathbb{H}^d\),
\begin{equation}\label{eq:Sw-strong-convexity}
S_w(\gamma(t))
\le
(1-t)S_w(\gamma(0))
+tS_w(\gamma(1))
-Wt(1-t)d^2\bigl(\gamma(0),\gamma(1)\bigr),
\qquad t\in[0,1].
\end{equation}
In particular, \(S_w\) is strictly geodesically convex on \(\mathbb{H}^d\).
\end{theorem}

\begin{proof}
By \eqref{eq:bg_strong_convexity}, for each \(i=1,\ldots,n\),
\[
d^2\!\bigl(x_i,\gamma(t)\bigr)
\le
(1-t)d^2\!\bigl(x_i,\gamma(0)\bigr)
+t d^2\!\bigl(x_i,\gamma(1)\bigr)
-t(1-t)d^2\!\bigl(\gamma(0),\gamma(1)\bigr).
\]
Multiplying by \(w_i\) and summing over \(i\) gives
\[
\sum_{i=1}^n w_i d^2\!\bigl(x_i,\gamma(t)\bigr)
\le
(1-t)\sum_{i=1}^n w_i d^2\!\bigl(x_i,\gamma(0)\bigr)
+t\sum_{i=1}^n w_i d^2\!\bigl(x_i,\gamma(1)\bigr)
-t(1-t)\Bigl(\sum_{i=1}^n w_i\Bigr)d^2\!\bigl(\gamma(0),\gamma(1)\bigr).
\]
Using the definition of \(S_w\) and \(W=\sum_{i=1}^n w_i\) yields \eqref{eq:Sw-strong-convexity}. Since \(W>0\), the final term is strictly negative whenever \(t\in(0,1)\) and \(\gamma(0)\neq \gamma(1)\), which proves strict geodesic convexity.
\end{proof}

The next corollary formalizes the fact that the weighted Fr\'echet mean on \(\mathbb{H}^d\) is uniquely defined. This is a standard consequence of Hadamard geometry, but we record it explicitly because it is used repeatedly in the sequel.

\begin{corollary}[Existence and uniqueness of the weighted Fr\'echet mean]\label{cor:weighted-frechet-exists-unique}
The minimization problem
\begin{equation}\label{eq:weighted-frechet-problem}
\hat{\mu}_w
=
\argmin_{\mu\in\mathbb{H}^d} S_w(\mu)
\end{equation}
has a unique solution.
\end{corollary}

\begin{proof}
Uniqueness follows immediately from the strict geodesic convexity established in Theorem~\ref{thm:Sw-strong-convexity}. For existence, choose an index \(j\) such that \(w_j>0\). Then
\[
S_w(\mu)\ge w_j d^2(x_j,\mu).
\]
As \(d(o,\mu)\to\infty\), the triangle inequality implies \(d(x_j,\mu)\to\infty\), so \(S_w(\mu)\to\infty\). Hence \(S_w\) is coercive. Since \(\mathbb{H}^d\) is proper and \(S_w\) is continuous, the minimum is attained.
\end{proof}

We now turn to the weighted maximum likelihood estimator for the Riemannian Gaussian model. The location and scale parameters behave very differently: the location parameter is governed by the weighted barycenter problem, whereas the scale parameter is determined by a one-dimensional convex profile optimization.

\begin{theorem}[Existence and uniqueness of the weighted Riemannian Gaussian MLE]\label{thm:single-mle}
Let \(\ell_w\) be the weighted log-likelihood \eqref{eq:weighted-loglik}. Then the following statements hold.
\begin{enumerate}[label=(\roman*)]
    \item The weighted location MLE \(\hat{\mu}_w\) exists and is unique.
    \item If the positively weighted sample is not a singleton, equivalently if \(S_w(\hat{\mu}_w)>0\), then the unconstrained weighted scale MLE
    \[
    \hat{\beta}_w
    =
    \argmin_{\beta>0}
    \bigl\{
    \beta S_w(\hat{\mu}_w)+W A_d(\beta)
    \bigr\}
    \]
    exists and is unique.
    \item If the positively weighted sample is a singleton, then \(S_w(\hat{\mu}_w)=0\) and the unconstrained weighted likelihood is unbounded above in \(\beta\); in particular, no finite scale MLE exists.
    \item On any compact interval \([\underline{\beta},\overline{\beta}]\subset(0,\infty)\), the constrained scale MLE always exists and is unique.
\end{enumerate}
\end{theorem}

\begin{proof}
Part (i) is exactly Corollary~\ref{cor:weighted-frechet-exists-unique}.

For parts (ii)--(iv), write
\[
g(\beta)
=
\beta S_\star + W A_d(\beta),
\qquad
S_\star:=S_w(\hat{\mu}_w)\ge 0.
\]
By Proposition~\ref{prop:Ad-derivatives}, \(A_d\) is strictly convex on \((0,\infty)\), hence so is \(g\). We first examine the behavior at the boundary of \((0,\infty)\).

As \(\beta\downarrow 0\), the integrand in \eqref{eq:normalizer-polar} increases pointwise to \(\sinh^{d-1}(r)\), whose integral over \([0,\infty)\) diverges. Therefore \(Z_d(\beta)\to\infty\), so \(A_d(\beta)\to\infty\), and hence \(g(\beta)\to\infty\) as \(\beta\downarrow 0\).

Now suppose \(S_\star>0\). By Proposition~\ref{prop:Zd-asymptotic},
\[
A_d(\beta)
=
-\frac{d}{2}\log\beta+O(1),
\qquad \beta\to\infty,
\]
and therefore
\[
g(\beta)
=
\beta S_\star-\frac{Wd}{2}\log\beta+O(1)
\to\infty
\qquad\text{as }\beta\to\infty.
\]
Since \(g\) is strictly convex and tends to \(+\infty\) at both ends of \((0,\infty)\), it has a unique minimizer. This proves part (ii).

If \(S_\star=0\), then
\[
g(\beta)=W A_d(\beta)
=
-\frac{Wd}{2}\log\beta+O(1)
\to -\infty
\qquad\text{as }\beta\to\infty.
\]
Equivalently, the weighted log-likelihood diverges to \(+\infty\). Thus no finite scale maximizer exists, proving part (iii).

Finally, on a compact interval \([\underline{\beta},\overline{\beta}]\), the function \(g\) is continuous and strictly convex. Hence it attains its minimum at a unique point, proving part (iv).
\end{proof}

The profile equation for the scale parameter can be written in a moment-matching form.

\begin{corollary}[Moment-matching characterization of \(\hat{\beta}_w\)]\label{cor:beta-moment-match}
In the nondegenerate case \(S_w(\hat{\mu}_w)>0\), the unique unconstrained scale MLE is the unique solution of
\begin{equation}\label{eq:beta-moment-match}
m_{2,d}(\beta)=\frac{S_w(\hat{\mu}_w)}{W}.
\end{equation}
\end{corollary}

\begin{proof}
Differentiating the profile objective
\[
\beta S_w(\hat{\mu}_w)+W A_d(\beta)
\]
with respect to \(\beta\) gives
\[
S_w(\hat{\mu}_w)+W A_d'(\beta)=0.
\]
Using the definition \(m_{2,d}(\beta)=-A_d'(\beta)\) yields \eqref{eq:beta-moment-match}. Uniqueness follows from the strict monotonicity of \(m_{2,d}\) in Proposition~\ref{prop:Ad-derivatives}.
\end{proof}

The preceding results are the weighted analogues of the unweighted maximum-likelihood characterization studied by \cite{said_2018_GaussianDistributionsRiemannian} and, more generally, by \cite{chen_2024_RiemannianRadialDistributions}. In particular, the single-component Riemannian Gaussian estimator inherits the favorable geometric structure of barycenter-based inference on Hadamard manifolds.

\subsection{Existence of the constrained finite-mixture MLE}\label{subsec:mixture-existence}

We now consider the finite-mixture model on the constrained parameter space \(\Theta_c\). Since the unrestricted likelihood is singular by Proposition~\ref{prop:mixture-singularity}, the natural question is whether the constrained likelihood is well posed. The answer is affirmative.

We first verify that the location parameter set is compact.

\begin{lemma}[Compactness of the location parameter set]\label{lem:Cn-compact}
The closed geodesic convex hull \(C_n\) defined in \eqref{eq:Cn-def} is compact.
\end{lemma}

\begin{proof}
Let
\[
R=\max_{1\le i\le n} d(o,x_i).
\]
Then \(x_1,\ldots,x_n\in \overline{B}(o,R)\), the closed ball of radius \(R\) centered at \(o\). Because \(\mathbb{H}^d\) is a Hadamard manifold, closed metric balls are geodesically convex. Therefore every geodesic joining two points of \(\overline{B}(o,R)\) remains inside \(\overline{B}(o,R)\), and hence the geodesic convex hull of the sample is contained in \(\overline{B}(o,R)\). Since \(C_n\) is closed by definition and bounded, and \(\mathbb{H}^d\) is proper, it is compact.
\end{proof}

\begin{theorem}[Existence of the constrained mixture MLE]\label{thm:constrained-mle-exists}
The observed log-likelihood \eqref{eq:observed-loglik} attains its maximum on \(\Theta_c\).
\end{theorem}

\begin{proof}
By Lemma~\ref{lem:Cn-compact}, \(C_n\) is compact. The simplex \(\Delta_{K-1}\) is compact, and the scale box \([\underline{\beta},\overline{\beta}]^K\) is compact as well. Therefore \(\Theta_c\) is compact.

Now fix \(i\in\{1,\ldots,n\}\). The map
\[
\theta\mapsto
f(x_i;\theta)
=
\sum_{k=1}^K
\pi_k
\exp\!\bigl(
-A_d(\beta_k)-\beta_k d^2(x_i,\mu_k)
\bigr)
\]
is continuous on \(\Theta_c\), since each block is continuous. Moreover, \(f(x_i;\theta)>0\) for every \(\theta\in\Theta_c\), because each Riemannian Gaussian density is strictly positive on \(\mathbb{H}^d\) and at least one mixing weight is positive. Hence
\[
\ell(\theta)=\sum_{i=1}^n \log f(x_i;\theta)
\]
is continuous on \(\Theta_c\). By the Weierstrass theorem, \(\ell\) attains its maximum on \(\Theta_c\).
\end{proof}

Global uniqueness should not be expected for finite mixtures. Even when a maximizer is isolated modulo permutation, label switching produces \(K!\) equivalent maximizers, and further nonuniqueness may arise from genuine multimodality of the observed likelihood.

\subsection{EM monotonicity and fixed-point convergence}\label{subsec:em-theory}

We next turn to the exact EM algorithm. The first result is the standard monotonicity property. Although the argument is classical, it is useful to write it in the notation of the present model because the later convergence theorem builds directly on it.

\begin{theorem}[Monotonicity of exact EM]\label{thm:em-monotonicity}
Let \(\{\theta^{(t)}\}_{t\ge 0}\) be the sequence generated by Algorithm~\ref{alg:exact-em}. Then
\begin{equation}\label{eq:em-monotonicity}
\ell\bigl(\theta^{(t+1)}\bigr)\ge \ell\bigl(\theta^{(t)}\bigr)
\qquad\text{for all }t\ge 0.
\end{equation}
\end{theorem}

\begin{proof}
Fix \(t\), and abbreviate \(r_{ik}^{(t)}\) by \(r_{ik}\). For any parameter \(\theta\), define
\[
s_{ik}(\theta)
:=
\pi_k
\exp\bigl(-A_d(\beta_k)-\beta_k d^2(x_i,\mu_k)\bigr),
\qquad
f_i(\theta):=\sum_{k=1}^K s_{ik}(\theta).
\]
Then \(\ell(\theta)=\sum_{i=1}^n \log f_i(\theta)\). Since \(\sum_{k=1}^K r_{ik}=1\), Jensen's inequality gives
\[
\log f_i(\theta)
=
\log\sum_{k=1}^K r_{ik}\frac{s_{ik}(\theta)}{r_{ik}}
\ge
\sum_{k=1}^K r_{ik}\log\frac{s_{ik}(\theta)}{r_{ik}}.
\]
Summing over \(i\) yields
\begin{equation}\label{eq:em-lower-bound}
\ell(\theta)
\ge
\mathcal{Q}(\theta\mid\theta^{(t)})
-
\sum_{i=1}^n\sum_{k=1}^K r_{ik}^{(t)}\log r_{ik}^{(t)}.
\end{equation}
At \(\theta=\theta^{(t)}\), equality holds because the responsibilities are chosen as the exact posterior probabilities:
\[
r_{ik}^{(t)}
=
\frac{s_{ik}(\theta^{(t)})}{f_i(\theta^{(t)})}.
\]
Therefore
\[
\ell(\theta)-\ell\bigl(\theta^{(t)}\bigr)
\ge
\mathcal{Q}(\theta\mid\theta^{(t)})
-
\mathcal{Q}\bigl(\theta^{(t)}\mid\theta^{(t)}\bigr).
\]
The exact M-step chooses \(\theta^{(t+1)}\) maximizing \(\mathcal{Q}(\cdot\mid\theta^{(t)})\), so
\[
\mathcal{Q}\bigl(\theta^{(t+1)}\mid\theta^{(t)}\bigr)
\ge
\mathcal{Q}\bigl(\theta^{(t)}\mid\theta^{(t)}\bigr).
\]
Substituting \(\theta=\theta^{(t+1)}\) proves \eqref{eq:em-monotonicity}.
\end{proof}

To formulate a convergence statement, we isolate the regime in which the exact EM update acts as a continuous self-map on a compact set.

\begin{assumption}[Interior compactness of the EM sequence]\label{ass:interior-compactness}
The exact EM iterates remain in a compact subset \(\mathcal{K}\) of the interior of the constrained parameter space:
\[
\mathcal{K}\subset \Theta_c^\circ
:=
\bigl\{
\theta\in\Theta_c:\ \pi_k>0\ \text{for all }k
\bigr\}.
\]
A sufficient condition is that the mixing weights are bounded away from zero, that is, there exists \(\underline{\pi}>0\) such that \(\pi_k^{(t)}\ge \underline{\pi}\) for all \(k\) and all \(t\).
\end{assumption}

Under this assumption, the blockwise updates that appear in the exact M-step are continuous.

\begin{lemma}[Continuity of the weighted barycenter map]\label{lem:barycenter-map-continuity}
Fix data points \(x_1,\ldots,x_n\in\mathbb{H}^d\). For every nonzero weight vector \(a=(a_1,\ldots,a_n)\in[0,\infty)^n\setminus\{0\}\), let
\[
\mathcal{B}(a)
=
\argmin_{\mu\in\mathbb{H}^d}\sum_{i=1}^n a_i d^2(x_i,\mu).
\]
Then \(\mathcal{B}\) is continuous on \([0,\infty)^n\setminus\{0\}\).
\end{lemma}

\begin{proof}
Let \(a^{(m)}\to a\) in \([0,\infty)^n\setminus\{0\}\), and define
\[
F_m(\mu)
=
\sum_{i=1}^n a_i^{(m)}d^2(x_i,\mu),
\qquad
F(\mu)
=
\sum_{i=1}^n a_i d^2(x_i,\mu).
\]
By Corollary~\ref{cor:weighted-frechet-exists-unique}, each function has a unique minimizer.

We first show that the family \(\{F_m\}\cup\{F\}\) is uniformly coercive. Since \(a\neq 0\), the total mass \(\sum_i a_i\) is strictly positive. Because \(a^{(m)}\to a\), there exist \(m_0\) and \(c_0>0\) such that \(\sum_i a_i^{(m)}\ge c_0\) for all \(m\ge m_0\). If \(R=\max_i d(o,x_i)\), then the triangle inequality implies
\[
d(x_i,\mu)\ge d(o,\mu)-R
\]
for every \(i\). Therefore,
\[
F_m(\mu)\ge c_0(d(o,\mu)-R)^2
\qquad\text{for all }m\ge m_0.
\]
Hence all minimizers lie in a common compact ball.

Let \(\mu_m=\mathcal{B}(a^{(m)})\). Any subsequence of \(\{\mu_m\}\) has a further subsequence converging to some \(\bar{\mu}\). On the common compact ball containing all minimizers, the functions \(F_m\) converge uniformly to \(F\), because the coefficients converge and \(\mu\mapsto d^2(x_i,\mu)\) is continuous. For every fixed \(\mu\in\mathbb{H}^d\), the minimizing property of \(\mu_m\) gives
\[
F(\mu_m)
\le
F_m(\mu_m)+\varepsilon_m
\le
F_m(\mu)+\varepsilon_m
\le
F(\mu)+2\varepsilon_m,
\]
where \(\varepsilon_m\to 0\). Passing to the convergent subsequence yields \(F(\bar{\mu})\le F(\mu)\) for every \(\mu\), so \(\bar{\mu}\) minimizes \(F\). By uniqueness, \(\bar{\mu}=\mathcal{B}(a)\). Thus every convergent subsequence of \(\{\mu_m\}\) converges to \(\mathcal{B}(a)\), which proves \(\mu_m\to\mathcal{B}(a)\).
\end{proof}

\begin{lemma}[Continuity of the constrained scale map]\label{lem:scale-map-continuity}
For \(S\ge 0\) and \(W>0\), define
\[
\mathcal{S}(S,W)
=
\argmin_{\beta\in[\underline{\beta},\overline{\beta}]}
\bigl\{\beta S+W A_d(\beta)\bigr\}.
\]
Then \(\mathcal{S}\) is continuous on \([0,\infty)\times(0,\infty)\).
\end{lemma}

\begin{proof}
Let \((S_m,W_m)\to(S,W)\) with \(W_m,W>0\). Define
\[
g_m(\beta)=\beta S_m+W_mA_d(\beta),
\qquad
g(\beta)=\beta S+W A_d(\beta),
\qquad
\beta\in[\underline{\beta},\overline{\beta}].
\]
Since \(A_d\) is continuous on the compact interval \([\underline{\beta},\overline{\beta}]\), the convergence \((S_m,W_m)\to(S,W)\) implies \(g_m\to g\) uniformly on that interval. By Proposition~\ref{prop:Ad-derivatives}, each \(g_m\) and \(g\) is strictly convex, hence has a unique minimizer. The same argument used in the proof of Lemma~\ref{lem:barycenter-map-continuity} shows that any accumulation point of the minimizers of \(g_m\) must minimize \(g\). By uniqueness, the minimizers converge to the minimizer of \(g\). Hence \(\mathcal{S}\) is continuous.
\end{proof}

\begin{proposition}[Continuity of the exact EM map]\label{prop:em-map-continuity}
Under Assumption~\ref{ass:interior-compactness}, the exact EM update defines a continuous self-map
\[
\mathcal{T}_{\mathrm{EM}}:\mathcal{K}\to\mathcal{K}.
\]
\end{proposition}

\begin{proof}
Fix \(\theta\in\mathcal{K}\). Because every \(\pi_k>0\) on \(\mathcal{K}\) and every component density is continuous and strictly positive on \(\mathbb{H}^d\), the responsibility formula \eqref{eq:responsibilities} defines a continuous function of \(\theta\). Hence the effective weights
\[
W_k(\theta)=\sum_{i=1}^n r_{ik}(\theta)
\]
and the weight vectors \((r_{1k}(\theta),\ldots,r_{nk}(\theta))\) are continuous in \(\theta\). The mixing-weight update
\[
\pi_k^+(\theta)=\frac{W_k(\theta)}{n}
\]
is therefore continuous.

The location update
\[
\mu_k^+(\theta)
=
\mathcal{B}\bigl(r_{1k}(\theta),\ldots,r_{nk}(\theta)\bigr)
\]
is continuous by Lemma~\ref{lem:barycenter-map-continuity}. The scale update
\[
\beta_k^+(\theta)
=
\mathcal{S}\bigl(S_k(\theta),W_k(\theta)\bigr),
\qquad
S_k(\theta):=\sum_{i=1}^n r_{ik}(\theta)d^2\bigl(x_i,\mu_k^+(\theta)\bigr),
\]
is continuous by Lemma~\ref{lem:scale-map-continuity}. Combining these blocks proves continuity of \(\mathcal{T}_{\mathrm{EM}}\). By Assumption~\ref{ass:interior-compactness}, the exact EM iterates remain in \(\mathcal{K}\), so \(\mathcal{T}_{\mathrm{EM}}\) acts as a self-map on \(\mathcal{K}\).
\end{proof}

\begin{theorem}[Fixed-point convergence of accumulation points for exact EM]\label{thm:em-convergence}
Under Assumption~\ref{ass:interior-compactness}, the exact EM sequence generated by Algorithm~\ref{alg:exact-em} satisfies the following properties.
\begin{enumerate}[label=(\roman*)]
    \item The observed log-likelihood sequence \(\ell(\theta^{(t)})\) is nondecreasing and converges to a finite limit \(\ell_\infty\).
    \item The sequence \(\{\theta^{(t)}\}\) has accumulation points in \(\mathcal{K}\).
    \item Every accumulation point \(\theta^\star\) is a fixed point of the exact EM map:
    \begin{equation}\label{eq:em-fixedpoint}
    \mathcal{T}_{\mathrm{EM}}(\theta^\star)=\theta^\star.
    \end{equation}
    \item Consequently, every accumulation point satisfies the exact mixture self-consistency equations
    \begin{align}
    \pi_k^\star &= \frac1n\sum_{i=1}^n r_{ik}(\theta^\star),\label{eq:self-consistency-pi}\\
    \mu_k^\star &= \arg\min_{\mu\in C_n}\sum_{i=1}^n r_{ik}(\theta^\star)d^2(x_i,\mu),\label{eq:self-consistency-mu}\\
    \beta_k^\star
    &=
    \arg\min_{\beta\in[\underline{\beta},\overline{\beta}]}
    \Biggl\{
    \beta\sum_{i=1}^n r_{ik}(\theta^\star)d^2(x_i,\mu_k^\star)
    +
    \Bigl(\sum_{i=1}^n r_{ik}(\theta^\star)\Bigr)A_d(\beta)
    \Biggr\}.
    \label{eq:self-consistency-beta}
    \end{align}
\end{enumerate}
\end{theorem}

\begin{proof}
Part (i) follows from Theorem~\ref{thm:em-monotonicity}. Since \(\ell\) is continuous on the compact set \(\mathcal{K}\), it is bounded above there, and therefore the monotone sequence \(\ell(\theta^{(t)})\) converges to a finite limit \(\ell_\infty\). Part (ii) follows from compactness of \(\mathcal{K}\).

For part (iii), define the improvement function
\[
D(\theta)
=
\mathcal{Q}\bigl(\mathcal{T}_{\mathrm{EM}}(\theta)\mid\theta\bigr)
-
\mathcal{Q}(\theta\mid\theta).
\]
By construction of the exact M-step, \(D(\theta)\ge 0\) for every \(\theta\in\mathcal{K}\). By Proposition~\ref{prop:em-map-continuity}, the map \(\mathcal{T}_{\mathrm{EM}}\) is continuous, hence \(D\) is continuous on \(\mathcal{K}\).

Let \(\theta^\star\) be an accumulation point, and choose a subsequence \(\theta^{(t_m)}\to\theta^\star\). Suppose, for contradiction, that \(\mathcal{T}_{\mathrm{EM}}(\theta^\star)\neq \theta^\star\). Because the exact M-step uniquely maximizes each block of \(\mathcal{Q}(\cdot\mid\theta^\star)\)---the simplex block is unique, the barycenter block is unique by Corollary~\ref{cor:weighted-frechet-exists-unique}, and the scale block is unique by strict convexity---we must have
\[
D(\theta^\star)>0.
\]
By continuity, there exist \(\varepsilon>0\) and \(m_0\) such that \(D(\theta^{(t_m)})\ge \varepsilon\) for all \(m\ge m_0\). On the other hand, the lower bound \eqref{eq:em-lower-bound} implies
\[
\ell\bigl(\theta^{(t_m+1)}\bigr)-\ell\bigl(\theta^{(t_m)}\bigr)
\ge
\mathcal{Q}\bigl(\theta^{(t_m+1)}\mid\theta^{(t_m)}\bigr)
-
\mathcal{Q}\bigl(\theta^{(t_m)}\mid\theta^{(t_m)}\bigr)
=
D\bigl(\theta^{(t_m)}\bigr)
\ge
\varepsilon.
\]
This contradicts the convergence of \(\ell(\theta^{(t)})\). Therefore \(\mathcal{T}_{\mathrm{EM}}(\theta^\star)=\theta^\star\).

Part (iv) is simply the blockwise form of the fixed-point equation. If \(\mathcal{T}_{\mathrm{EM}}(\theta^\star)=\theta^\star\), then the updated mixing weights, locations, and scales equal their current values, which is precisely the system \eqref{eq:self-consistency-pi}--\eqref{eq:self-consistency-beta}.
\end{proof}

The fixed-point formulation above is the most natural one for the present manuscript, because it is intrinsic to the EM update map and avoids introducing local coordinates on the simplex or on \(\mathbb{H}^d\). In smooth interior parameterizations, exact EM fixed points coincide with stationary points of the observed log-likelihood, but the fixed-point description already captures the mode of convergence most relevant for mixture computation.

Assumption~\ref{ass:interior-compactness} is also mild from a practical standpoint. Its purpose is to exclude boundary pathologies, such as vanishing mixture weights, that are not specific to hyperbolic geometry. In implementation, such issues can be handled by reinitializing nearly empty components or by imposing a small lower bound on the mixing weights.

\subsection{Generalized EM and monotonicity under truncated inner solves}\label{subsec:gem-theory}

We now show that a truncated inner barycenter update still yields likelihood ascent as long as each component variance is not increased. This is the key theoretical justification for the GEM scheme introduced in Section~\ref{sec:computation}.

\begin{proposition}[Monotonicity of generalized EM with descent inner steps]\label{prop:gem-monotone}
Consider one outer EM iteration. Suppose that, for each component \(k\), the location update produces a point \(\widetilde{\mu}_k\in C_n\) satisfying
\begin{equation}\label{eq:descent-inner}
S_k^{(t)}(\widetilde{\mu}_k)\le S_k^{(t)}\bigl(\mu_k^{(t)}\bigr).
\end{equation}
If the mixing weights are then updated exactly by \eqref{eq:pi-update} and the scales are updated exactly by \eqref{eq:beta-update}, the resulting generalized EM iterate \(\widetilde{\theta}^{(t+1)}\) satisfies
\begin{equation}\label{eq:gem-monotonicity}
\ell\bigl(\widetilde{\theta}^{(t+1)}\bigr)\ge \ell\bigl(\theta^{(t)}\bigr).
\end{equation}
\end{proposition}

\begin{proof}
By \eqref{eq:Q-block-structure},
\[
\mathcal{Q}(\theta\mid\theta^{(t)})
=
\sum_{k=1}^K
\Big[
W_k^{(t)}\log\pi_k
-
W_k^{(t)}A_d(\beta_k)
-
\beta_k S_k^{(t)}(\mu_k)
\Big].
\]
If \eqref{eq:descent-inner} holds, then for fixed \(\beta_k\) the \(k\)th location block does not decrease the corresponding contribution to \(\mathcal{Q}\). The subsequent exact updates of \(\pi_k\) and \(\beta_k\) maximize the remaining blocks and therefore also do not decrease \(\mathcal{Q}\). Hence
\[
\mathcal{Q}\bigl(\widetilde{\theta}^{(t+1)}\mid\theta^{(t)}\bigr)
\ge
\mathcal{Q}\bigl(\theta^{(t)}\mid\theta^{(t)}\bigr).
\]
Applying the same lower-bound argument as in the proof of Theorem~\ref{thm:em-monotonicity} yields \eqref{eq:gem-monotonicity}.
\end{proof}

\begin{corollary}[Truncated MM inner loops define a valid GEM scheme]\label{cor:truncated-mm-gem}
If the inner location updates in Algorithm~\ref{alg:gem} are implemented by any finite number \(L\ge 1\) of MM steps from Algorithm~\ref{alg:mm-barycenter}, then the resulting generalized EM algorithm is monotone.
\end{corollary}

\begin{proof}
By Theorem~\ref{thm:mm-convergence}, every MM step decreases the weighted Fr\'echet objective. Therefore, after any finite number \(L\ge 1\) of inner MM steps, condition \eqref{eq:descent-inner} holds for each component. The conclusion follows immediately from Proposition~\ref{prop:gem-monotone}.
\end{proof}

\subsection{Computational complexity and practical interpretation}\label{subsec:complexity}

The preceding theory also clarifies the computational profile of the proposed algorithms. For fixed \(K\) and dimension \(d\), the E-step of Algorithm~\ref{alg:exact-em} costs \(O(nKd)\) once hyperbolic distances are evaluated through Lorentzian inner products. Each inner MM step for a component mean costs \(O(nd)\), so if \(L_k\) inner steps are used for component \(k\), the total location-update cost is \(O\!\bigl(nd\sum_{k=1}^K L_k\bigr)\). By contrast, the scale updates are one dimensional and become essentially negligible once \(A_d\), \(A_d'\), and \(A_d''\) have been precomputed.

Finally, it is worth emphasizing once more why the weighted formulation is central. The weighted single-component problem is not merely a technical variation of ordinary maximum likelihood estimation; it is the exact M-step subproblem in finite mixtures, and it is the natural primitive for further extensions such as classification EM, penalized component-wise estimation, and Bayesian or nonparametric formulations in which posterior weights replace hard sample counts. For this reason, the weighted log-likelihood \eqref{eq:weighted-loglik} should be regarded as the primary inferential object, with the unweighted MLE as a special case.

%%%%%%%%%%%%%%%%%%%%%%%%%%%%%%%%%%%%%%%%%%%%%%%%%%%%%%%%%%%%%%%%%%%%%%%%%%%%%%%%%
%%%%%%%%%%%%%%%%%%%%%%%%%%%%%%%%%%%%%%%%%%%%%%%%%%%%%%%%%%%%%%%%%%%%%%%%%%%%%%%%%
\section{Experiments}\label{sec:experiment}

\subsection{Simulation 1: weighted single-component estimation}\label{subsec:sim1}

We first examine the weighted single-component estimator, since this is the basic primitive that later reappears in the M-step of the finite-mixture model. The purpose of this experiment is therefore twofold. First, it provides a direct empirical validation of the weighted maximum likelihood estimator for a single Riemannian Gaussian distribution. Second, it isolates the numerical behavior of the weighted barycenter and scale profile updates before they are embedded inside the more complicated mixture algorithm.

In this initial study, we focus on the hyperbolic plane \(\mathbb{H}^2\subset\mathbb{R}^3\) and generate data from a single Riemannian Gaussian distribution with inverse-scale parameter \(\beta_0=2\) and center
\[
\mu_0=
\left(
\frac{\sinh(1.5)}{\sqrt{2}},
\frac{\sinh(1.5)}{\sqrt{2}},
\cosh(1.5)
\right)\in\mathbb{H}^2.
\]
This choice places the center away from the origin and avoids any artificial symmetry that might simplify the estimation problem. For each sample size $n \in \lbrace 100, 200, \ldots, 1000\rbrace$,  we generate \(100\) independent Monte Carlo replicates. To emphasize the weighted nature of the estimator, we assign observation-specific positive weights \(w_1,\ldots,w_n\) independently from a Gamma distribution and rescale them to satisfy \(\sum_{i=1}^n w_i=n\). We consider two regimes of weight heterogeneity: a mild regime with shape parameter \(a=5\), and a strong regime with \(a=0.5\). For each replicate, we fit the weighted Riemannian Gaussian model and record the geodesic location error \(d(\hat\mu_w,\mu_0)\), the relative scale error \(|\hat\beta_w-\beta_0|/\beta_0\), the weighted score residual
\[
\left\|
\sum_{i=1}^n w_i \operatorname{Log}_{\hat\mu_w}(x_i)
\right\|_L,
\]
the runtime, and optimization diagnostics from the barycenter solver.

\begin{figure}[t]
\centering
\begin{subfigure}{0.49\textwidth}
    \centering
    \includegraphics[width=\textwidth]{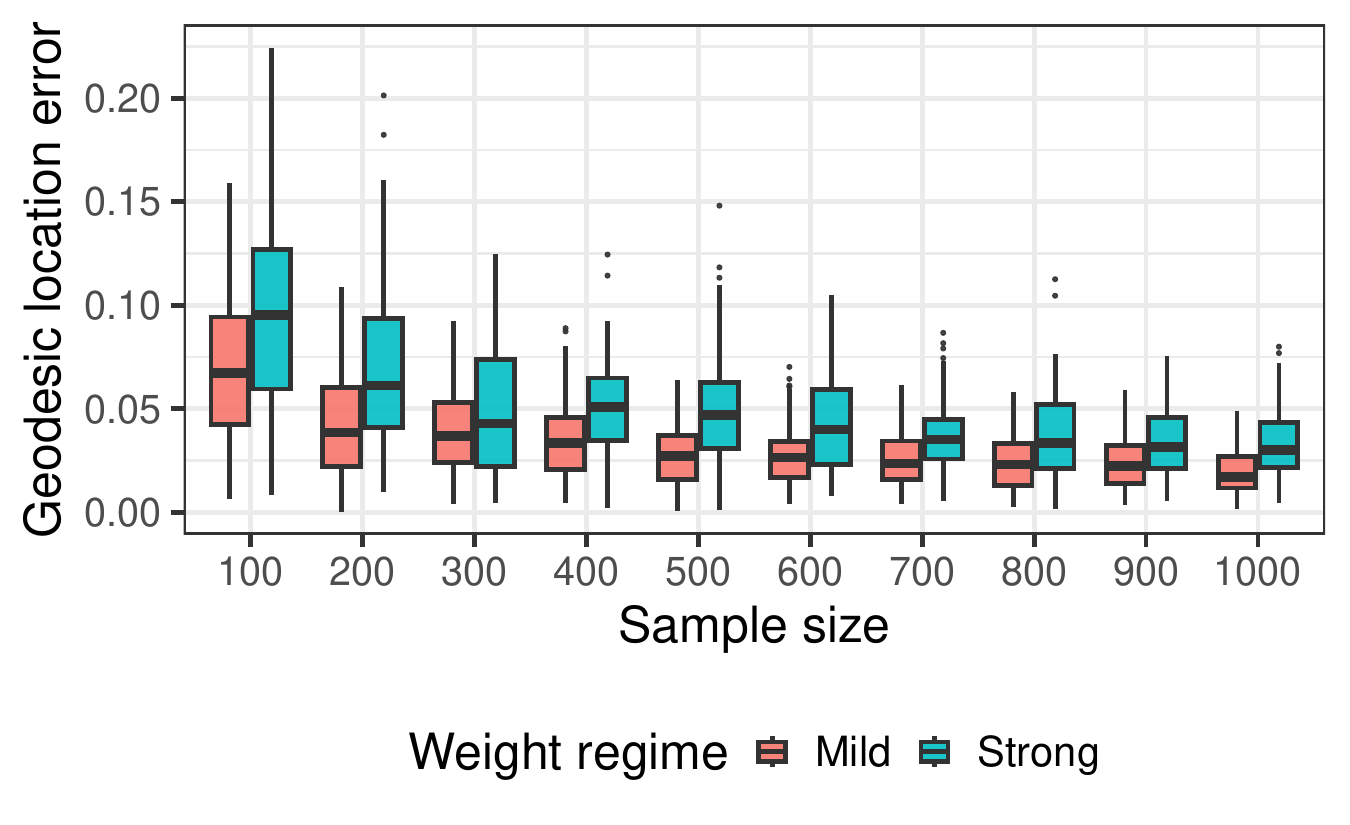}
    \caption{Geodesic location error.}
\end{subfigure}
\hfill
\begin{subfigure}{0.49\textwidth}
    \centering
    \includegraphics[width=\textwidth]{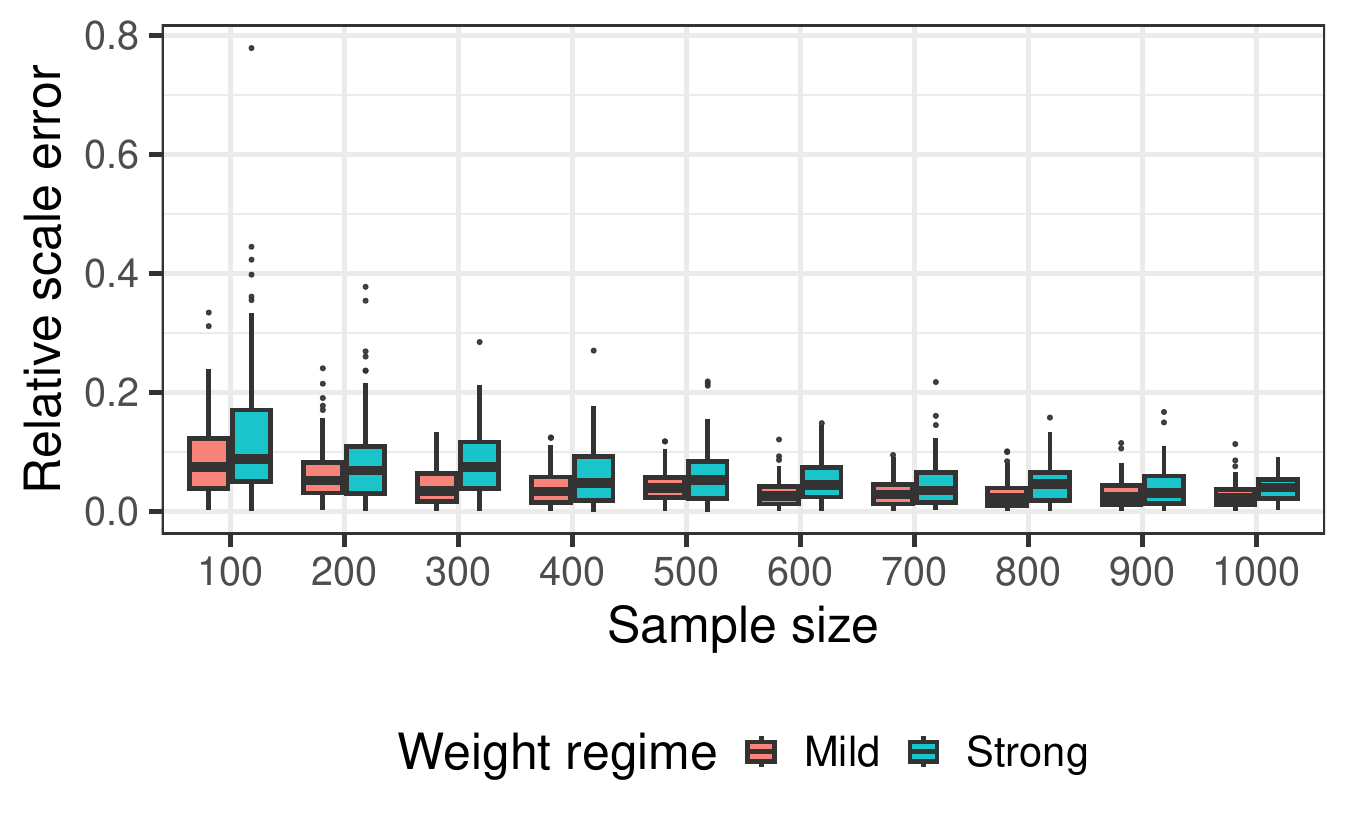}
    \caption{Relative scale error.}
\end{subfigure}
\caption{Estimation accuracy for weighted single-component estimation. Boxplots show geodesic location error and relative inverse-scale error across Monte Carlo replicates; colors distinguish mild and strong weight heterogeneity.under mild and strong weight heterogeneity.}
\label{fig:sim1-location-scale}
\end{figure}

\begin{figure}[t]
\centering
\begin{subfigure}{0.49\textwidth}
    \centering
    \includegraphics[width=\textwidth]{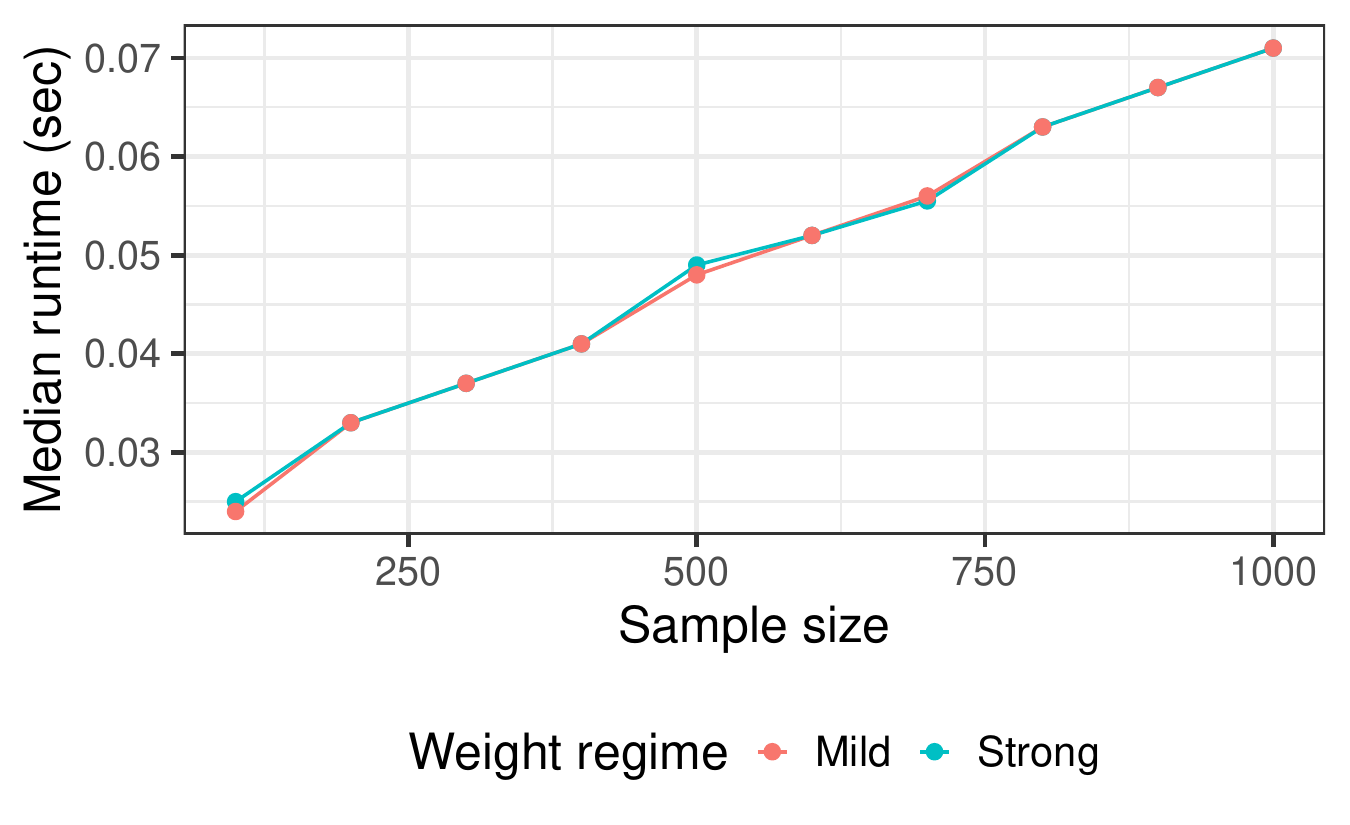}
    \caption{Median runtime.}
\end{subfigure}
\hfill
\begin{subfigure}{0.49\textwidth}
    \centering
    \includegraphics[width=\textwidth]{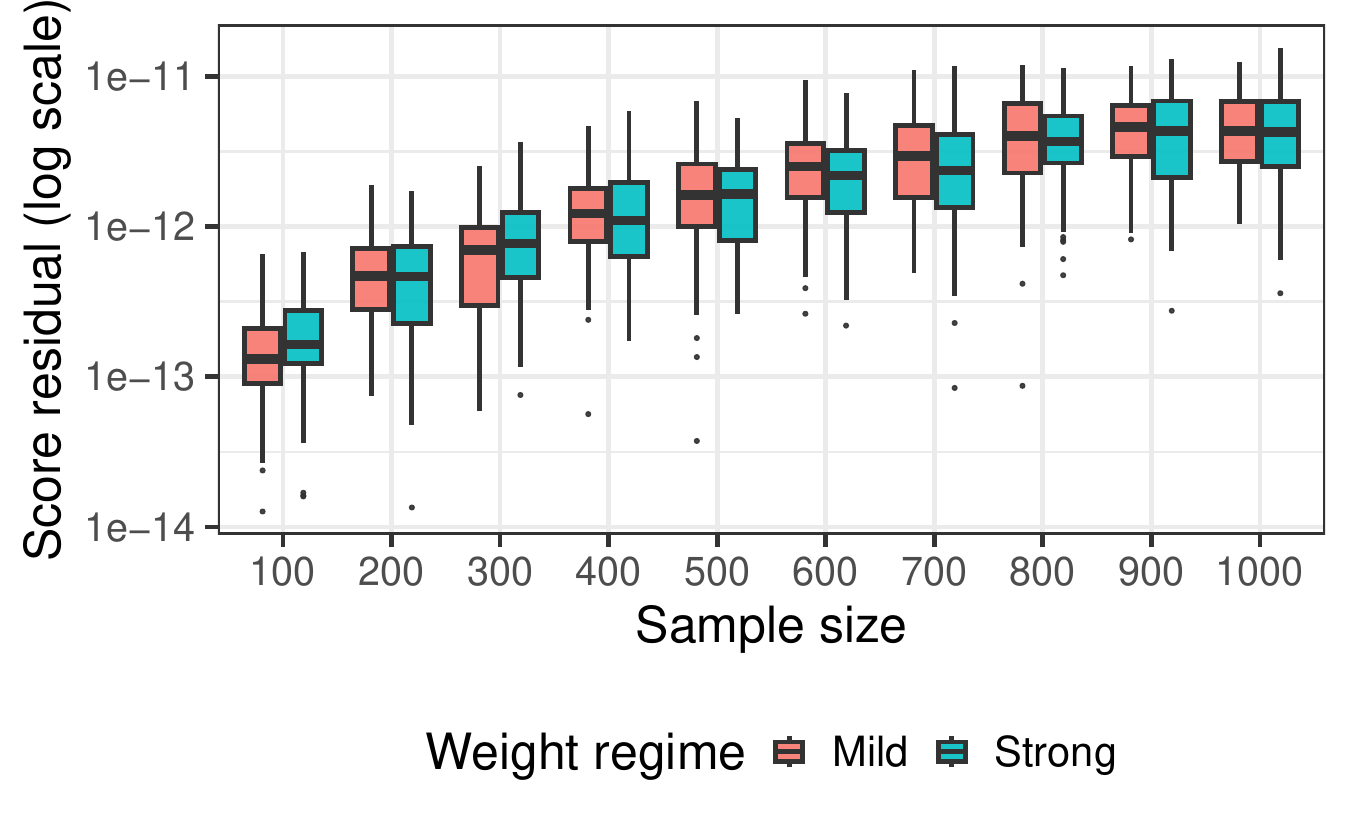}
    \caption{Weighted score residuals.}
\end{subfigure}
\caption{
Computational diagnostics for weighted single-component estimation. 
The left panel reports median runtime as a function of sample size, and the right panel shows weighted score residuals on a logarithmic scale.
}
\label{fig:sim1-runtime-score}
\end{figure}

Table~\ref{tab:sim1-main} reports the main numerical summaries, using medians and interquartile ranges across Monte Carlo replicates. Figures~\ref{fig:sim1-location-scale} and \ref{fig:sim1-runtime-score} visualize the corresponding distributions and runtime behavior.

\begin{table}[t]
\centering
\caption{
Weighted single-component estimation on \(\mathbb{H}^2\). 
Reported values are medians with interquartile ranges in parentheses over \(100\) Monte Carlo replicates.
}
\label{tab:sim1-main}
% Simulation 1 summary.
% tab:sim1-main
\begin{tabular}{llcccc}
\hline
Weight regime & n & Location error & Relative scale error & Score residual & Runtime (sec) \\
\hline
\multirow{10}{*}{Mild} & 100 & 0.067 (0.052) & 0.075 (0.084) & 1.16e-13 & 0.024 (0.005) \\
 & 200 & 0.039 (0.038) & 0.052 (0.052) & 3.91e-13 & 0.033 (0.002) \\
 & 300 & 0.037 (0.029) & 0.034 (0.048) & 5.69e-13 & 0.037 (0.001) \\
 & 400 & 0.033 (0.025) & 0.034 (0.043) & 1.12e-12 & 0.041 (0.003) \\
 & 500 & 0.027 (0.021) & 0.040 (0.034) & 1.49e-12 & 0.048 (0.004) \\
 & 600 & 0.026 (0.017) & 0.026 (0.029) & 2.38e-12 & 0.052 (0.002) \\
 & 700 & 0.024 (0.019) & 0.029 (0.033) & 2.70e-12 & 0.056 (0.004) \\
 & 800 & 0.023 (0.020) & 0.023 (0.029) & 3.89e-12 & 0.063 (0.005) \\
 & 900 & 0.022 (0.018) & 0.025 (0.032) & 4.43e-12 & 0.067 (0.002) \\
 & 1000 & 0.017 (0.015) & 0.024 (0.025) & 4.15e-12 & 0.071 (0.006) \\
\hline
\multirow{10}{*}{Strong} & 100 & 0.095 (0.067) & 0.088 (0.120) & 1.56e-13 & 0.025 (0.005) \\
 & 200 & 0.061 (0.053) & 0.069 (0.079) & 3.34e-13 & 0.033 (0.001) \\
 & 300 & 0.043 (0.051) & 0.075 (0.078) & 6.72e-13 & 0.037 (0.001) \\
 & 400 & 0.051 (0.030) & 0.048 (0.074) & 9.99e-13 & 0.041 (0.004) \\
 & 500 & 0.047 (0.032) & 0.054 (0.062) & 1.57e-12 & 0.049 (0.004) \\
 & 600 & 0.040 (0.036) & 0.045 (0.049) & 2.07e-12 & 0.052 (0.002) \\
 & 700 & 0.035 (0.019) & 0.035 (0.051) & 2.23e-12 & 0.055 (0.004) \\
 & 800 & 0.034 (0.031) & 0.047 (0.047) & 3.66e-12 & 0.063 (0.005) \\
 & 900 & 0.032 (0.025) & 0.032 (0.047) & 4.10e-12 & 0.067 (0.002) \\
 & 1000 & 0.030 (0.022) & 0.040 (0.032) & 4.23e-12 & 0.071 (0.005) \\
\hline
\end{tabular}

\end{table}

Several features are clear from these results. First, both the location and scale estimators improve steadily as the sample size increases. Under mild weight heterogeneity, the median geodesic location error decreases from \(0.067\) at \(n=100\) to \(0.017\) at \(n=1000\), while the median relative scale error decreases from \(0.075\) to \(0.024\). Under strong weight heterogeneity, estimation is uniformly more difficult, as expected, but the same overall pattern remains: the median location error decreases from \(0.095\) at \(n=100\) to \(0.030\) at \(n=1000\), and the median relative scale error decreases from \(0.088\) to \(0.040\). Thus the weighted estimator remains accurate even when the observation weights are highly uneven, although the stronger heterogeneity induces noticeably larger dispersion, especially at smaller sample sizes.

Second, the computational cost is modest and grows smoothly with the sample size. The median runtime increases from roughly \(0.024\)–\(0.025\) seconds at \(n=100\) to about \(0.071\) seconds at \(n=1000\), with essentially no difference between the two weight regimes; see Figure~\ref{fig:sim1-runtime-score}(a). This is consistent with the fact that the dominant operations in the algorithm are linear in \(n\), and it suggests that the weighted single-component fit is not itself a computational bottleneck.

Third, the weighted score residuals are uniformly tiny across all settings. In median, they range from approximately \(1.16\times 10^{-13}\) to \(4.43\times 10^{-12}\) under mild heterogeneity and from \(1.56\times 10^{-13}\) to \(4.23\times 10^{-12}\) under strong heterogeneity. The slight increase with sample size visible in Figure~\ref{fig:sim1-runtime-score}(b) is negligible on an absolute scale and is best interpreted as accumulated floating-point noise rather than a deterioration in numerical quality. In particular, these residuals indicate that the computed solutions satisfy the weighted first-order equation to near machine precision.

Finally, no failed fits were observed in any of the \(2\times 10\times 100=2000\) Monte Carlo runs. 
The uniformly small score residuals indicate that the weighted first-order equation is satisfied to near machine precision, and the runtime results show that the weighted single-component fit remains computationally inexpensive over the sample sizes considered.

Overall, this example provides strong support for the weighted single-component Riemannian Gaussian estimator. The estimator is statistically accurate, numerically stable, and computationally inexpensive, and its performance degrades gracefully under substantial weight heterogeneity. Since this weighted problem is precisely the atomic subproblem that appears in the M-step of the finite-mixture model, these results provide direct empirical justification for building the mixture algorithm on top of the weighted formulation developed in Section~\ref{sec:model}.

\subsection{Simulation 2: four-component finite mixtures on \(\mathbb{H}^2\)}\label{subsec:sim2}

We next examine the full finite-mixture procedure in a symmetric four-component configuration on the hyperbolic plane. Let \(\rho=2\), and define $a=\frac{\sinh(\rho)}{\sqrt{2}}$ and $c = \cosh(\rho)$.
The four component centers are placed at
\[
\mu_1=(a,a,c),\qquad
\mu_2=(-a,a,c),\qquad
\mu_3=(-a,-a,c),\qquad
\mu_4=(a,-a,c),
\]
so that they lie symmetrically in the four quadrants of the \((x,y)\)-plane in the hyperboloid model. We use equal mixing proportions $\pi_1=\pi_2=\pi_3=\pi_4=\tfrac{1}{4}$ and a common inverse-scale parameter \(\beta_1=\cdots=\beta_4=1.5\). This yields a well-separated but nontrivial clustering problem. Although the implementation allows higher-dimensional extensions by zero-padding additional coordinates, we report only the \(\mathbb{H}^2\) case here. The simulation consists of three parts: mixture recovery, model selection, and a computational comparison between exact EM and GEM. We use \(10\) Monte Carlo replicates for mixture recovery and the EM/GEM comparison, and \(5\) replicates for model selection.

\begin{figure}[t]
\centering
\begin{subfigure}{0.32\textwidth}
    \centering
    \includegraphics[width=\textwidth]{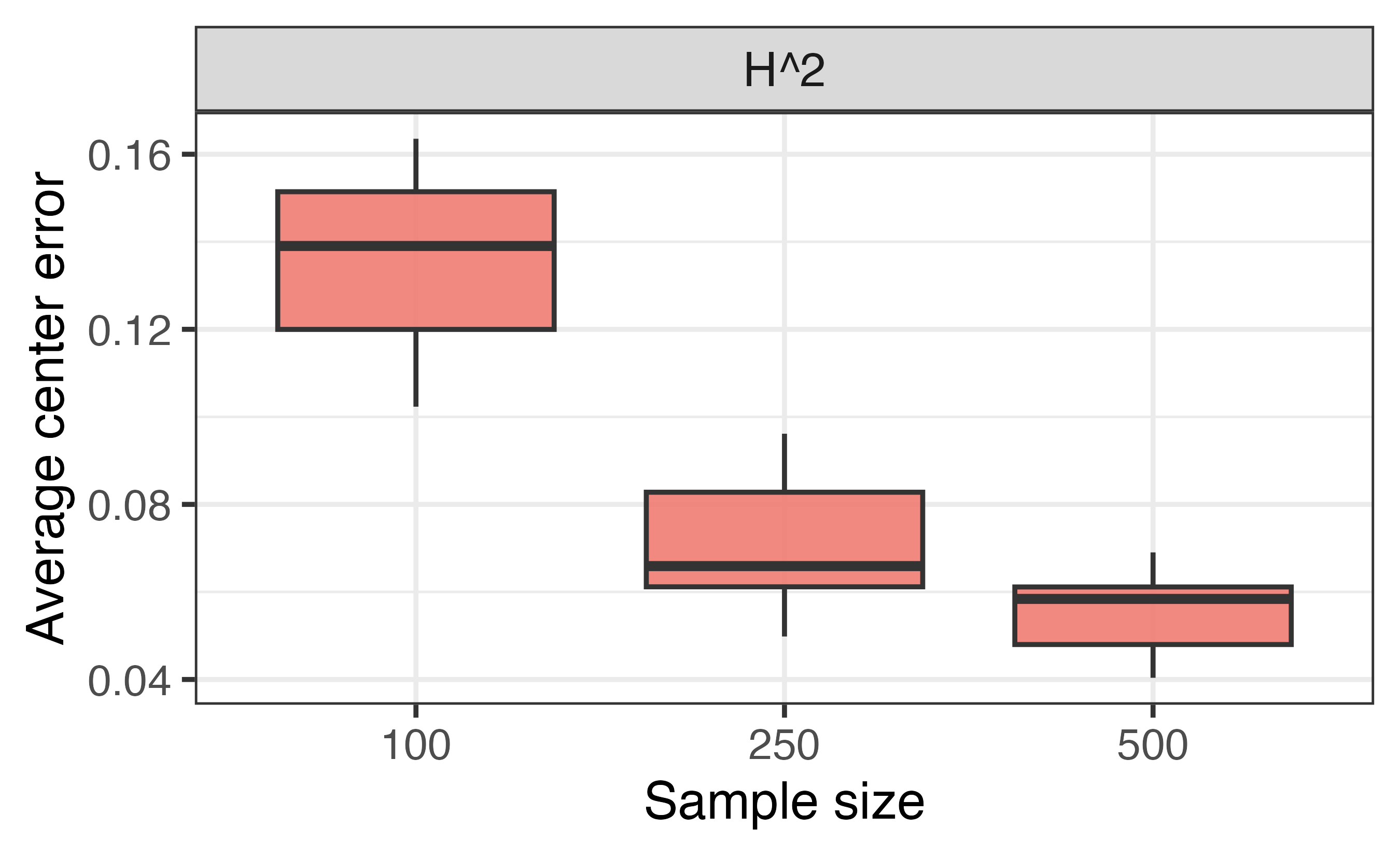}
    \caption{Average center error.}
\end{subfigure}
\hfill
\begin{subfigure}{0.32\textwidth}
    \centering
    \includegraphics[width=\textwidth]{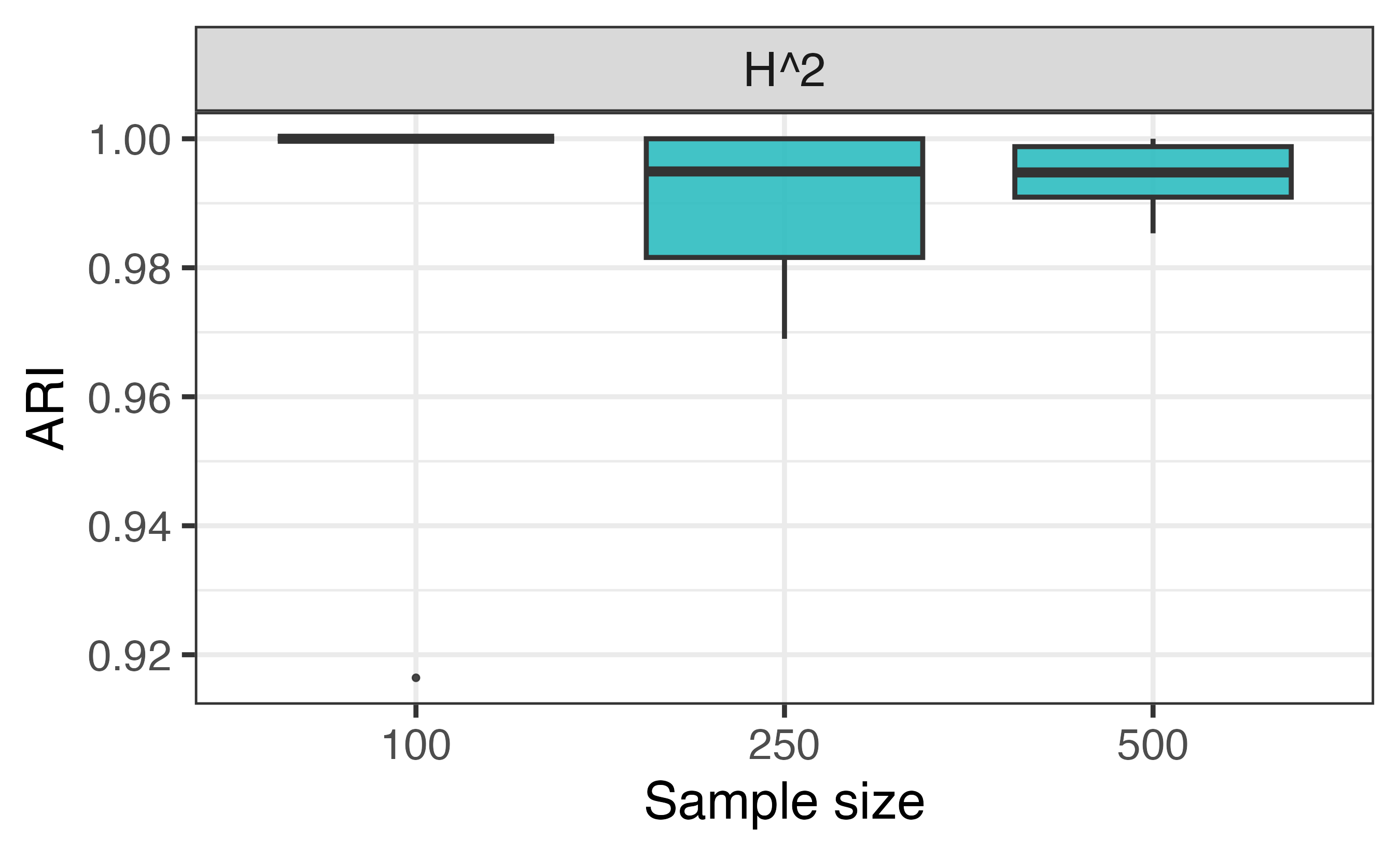}
    \caption{Adjusted Rand index.}
\end{subfigure}
\hfill
\begin{subfigure}{0.32\textwidth}
    \centering
    \includegraphics[width=\textwidth]{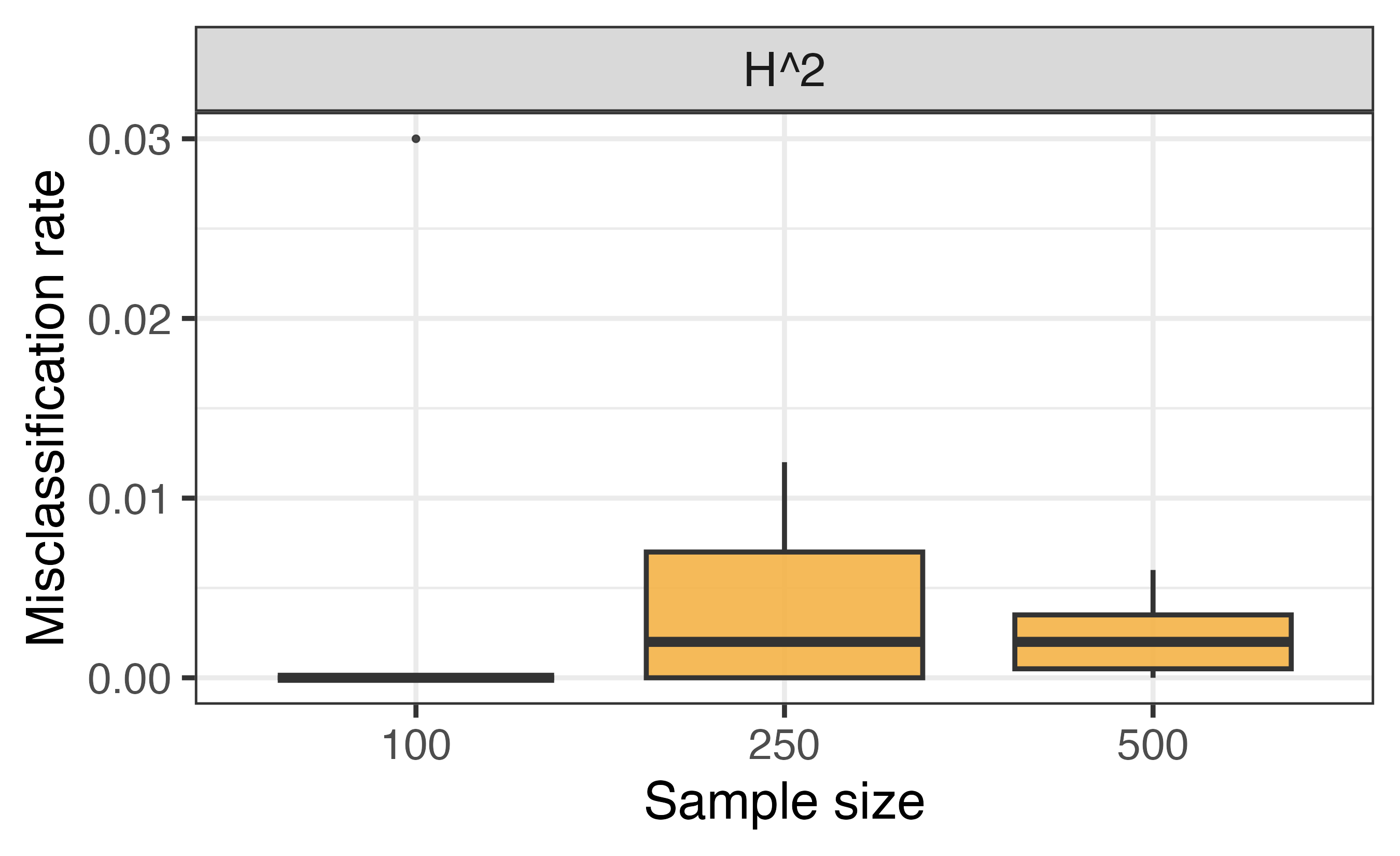}
    \caption{Misclassification rate.}
\end{subfigure}
\caption{
Recovery performance for the correctly specified four-component Riemannian Gaussian mixture. 
Boxplots show average geodesic center error, adjusted Rand index, and misclassification rate across Monte Carlo replicates.
}\label{fig:sim2-recovery}
\end{figure}

\paragraph{Mixture recovery.}
We first study the case in which the number of components is correctly specified as \(K=4\). Data are generated at sample sizes \(n\in\{100,250,500\}\), and for each replicate the fitted mixture is aligned to the true labeling by minimizing the total geodesic center error over all permutations of the component labels. Table~\ref{tab:sim2-recovery-main} summarizes the results, and Figure~\ref{fig:sim2-recovery} visualizes the distributions of the main performance metrics.

\begin{table}[t]
\centering
\small
\caption{
Finite-mixture recovery on \(\mathbb{H}^2\) with the number of components correctly specified as \(K=4\). 
Reported values are medians with interquartile ranges in parentheses.
}
\label{tab:sim2-recovery-main}
% Simulation 2A: mixture recovery summary.
% tab:sim2-recovery-main
\begin{tabular}{lllllll}
\hline
n & Center error & Rel. scale error & Mixing-weight error & ARI & Misclass. & Runtime (sec) \\
\hline
100 & 0.139 (0.031) & 0.166 (0.092) & 0.150 (0.105) & 1.000 (0.000) & 0.000 (0.000) & 12.547 (4.542) \\
250 & 0.066 (0.022) & 0.125 (0.050) & 0.075 (0.041) & 0.995 (0.018) & 0.002 (0.007) & 23.499 (5.156) \\
500 & 0.058 (0.013) & 0.067 (0.052) & 0.068 (0.037) & 0.995 (0.008) & 0.002 (0.003) & 39.422 (1.725) \\
\hline
\end{tabular}

\end{table}

Several features are clear. First, parameter recovery improves with the sample size. The median average center error decreases from \(0.139\) at \(n=100\) to \(0.066\) at \(n=250\) and further to \(0.058\) at \(n=500\). The median relative scale error exhibits the same trend, decreasing from \(0.166\) to \(0.125\) and then to \(0.067\). The \(\ell_1\)-error of the estimated mixing proportions also drops substantially, from \(0.150\) at \(n=100\) to \(0.068\) at \(n=500\). Second, clustering performance is already extremely strong at the smallest sample size and remains near perfect thereafter. The median ARI equals \(1.000\) at \(n=100\) and \(0.995\) at both \(n=250\) and \(n=500\), while the median misclassification rates are \(0.000\), \(0.002\), and \(0.002\), respectively. Third, all recovery runs converged successfully and no failures were observed. Runtime increases with the sample size, as expected, from a median of \(12.547\) seconds at \(n=100\) to \(39.422\) seconds at \(n=500\). Overall, this experiment shows that when the model is correctly specified, the proposed Riemannian Gaussian mixture estimator recovers both parameters and latent labels accurately.

\begin{figure}[t]
\centering
\begin{subfigure}{0.55\textwidth}
    \centering
    \includegraphics[width=\textwidth]{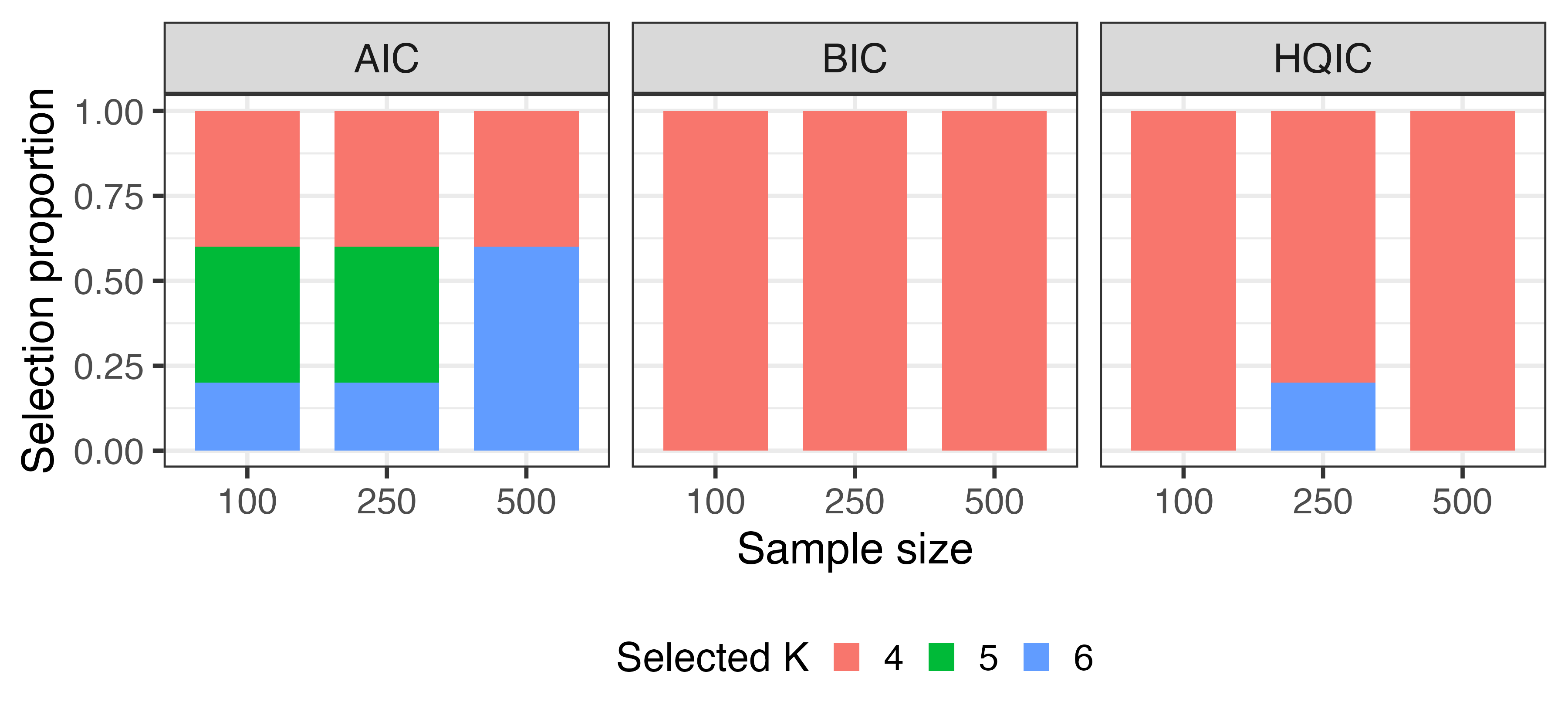}
    \caption{Selection frequencies.}
\end{subfigure}
\hfill
\begin{subfigure}{0.42\textwidth}
    \centering
    \includegraphics[width=\textwidth]{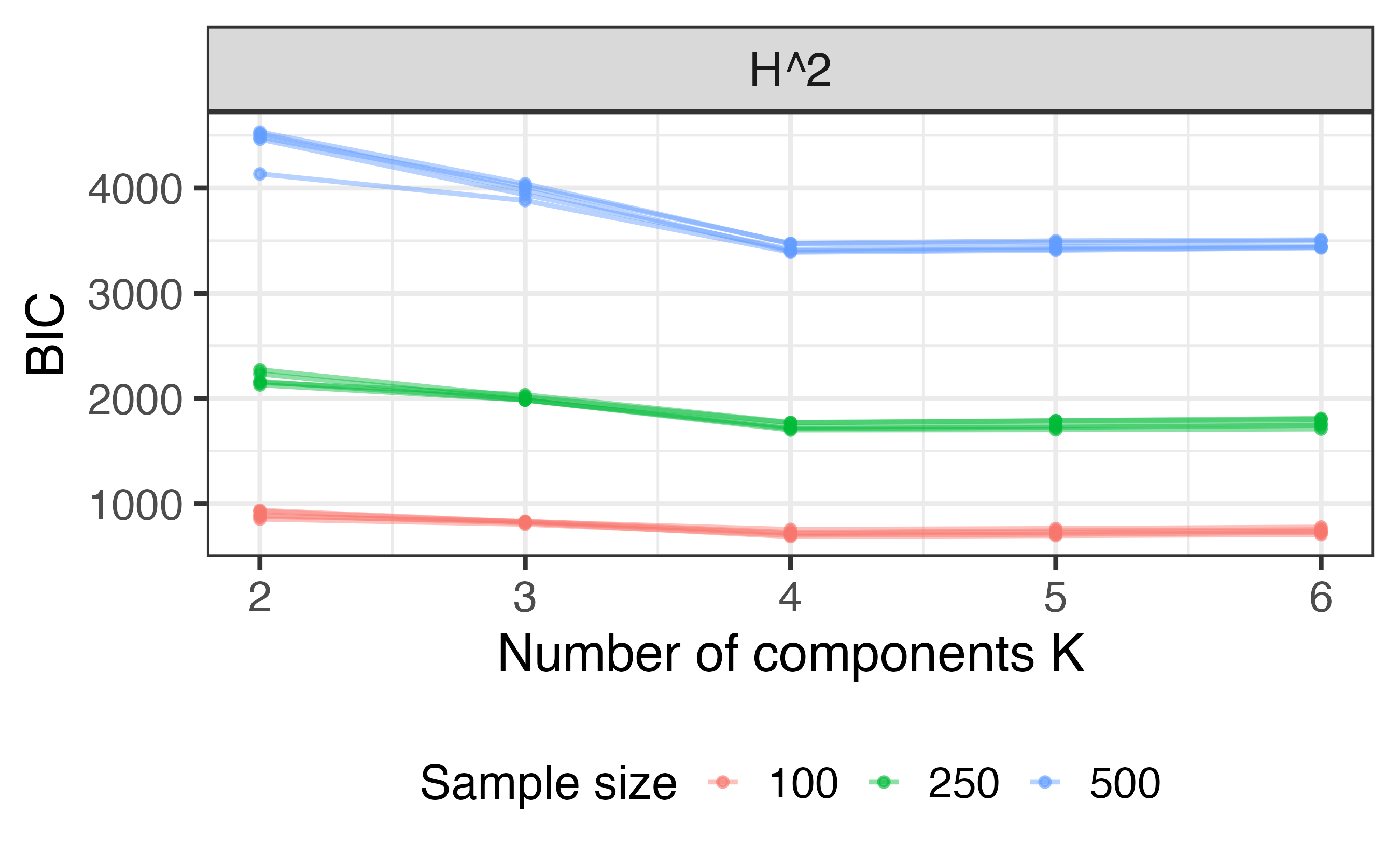}
    \caption{BIC paths over \(K\).}
\end{subfigure}
\caption{
Model-selection summaries for the four-component mixture experiment. 
The left panel shows the selected number of components under AIC, BIC, and HQIC, while the right panel shows BIC values over candidate \(K\), with lower values preferred.
}\label{fig:sim2-model-selection}
\end{figure}

\paragraph{Model selection.}
We next examine whether information criteria recover the correct number of components. For this experiment, we generate data from the same four-component mixture and fit models with \(K=2,\ldots,6\). The reduced range of candidate values is sufficient for distinguishing underfitting, correct specification, and moderate overfitting. Table~\ref{tab:sim2-model-selection-main} reports the empirical selection frequencies, and Figure~\ref{fig:sim2-model-selection} displays the same information graphically together with the BIC trajectories as a function of \(K\).

\begin{table}[t]
\centering
\small
\caption{
Empirical model-selection measures for  data generated from a four-component Riemannian Gaussian mixture.
}\label{tab:sim2-model-selection-main}
% Simulation 2B: empirical model-selection frequencies.
% tab:sim2-model-selection-main
\begin{tabular}{ccccc}
\hline
$n$ & Criterion & $K=4$ & $K=5$ & $K=6$ \\
\hline
100 & AIC & 0.40 & 0.40 & 0.20 \\
250 & AIC & 0.40 & 0.40 & 0.20 \\
500 & AIC & 0.40 & 0.00 & 0.60 \\
100 & BIC & 1.00 & 0.00 & 0.00 \\
250 & BIC & 1.00 & 0.00 & 0.00 \\
500 & BIC & 1.00 & 0.00 & 0.00 \\
100 & HQIC & 1.00 & 0.00 & 0.00 \\
250 & HQIC & 0.80 & 0.00 & 0.20 \\
500 & HQIC & 1.00 & 0.00 & 0.00 \\
\hline
\end{tabular}

\end{table}

The three criteria behave in a way that is fully consistent with classical finite-mixture practice. BIC selects the true value \(K=4\) in every run at all three sample sizes. HQIC is also highly reliable, selecting \(K=4\) in all runs except one at \(n=250\), where it selects \(K=6\) once. By contrast, AIC exhibits the familiar tendency to overfit: it selects \(K=4\) in only \(40\%\) of runs at each sample size and allocates the remaining mass to \(K=5\) or \(K=6\). The BIC path plot further shows that, across replicates and sample sizes, the criterion typically reaches its minimum around the true value \(K=4\). Thus, even in this reduced Monte Carlo setting, the model-selection behavior of the hyperbolic mixture model closely parallels that of classical Euclidean mixtures, with BIC providing the clearest recovery of the true number of components.

\begin{figure}[t]
\centering
\begin{subfigure}{0.32\textwidth}
    \centering
    \includegraphics[width=\textwidth]{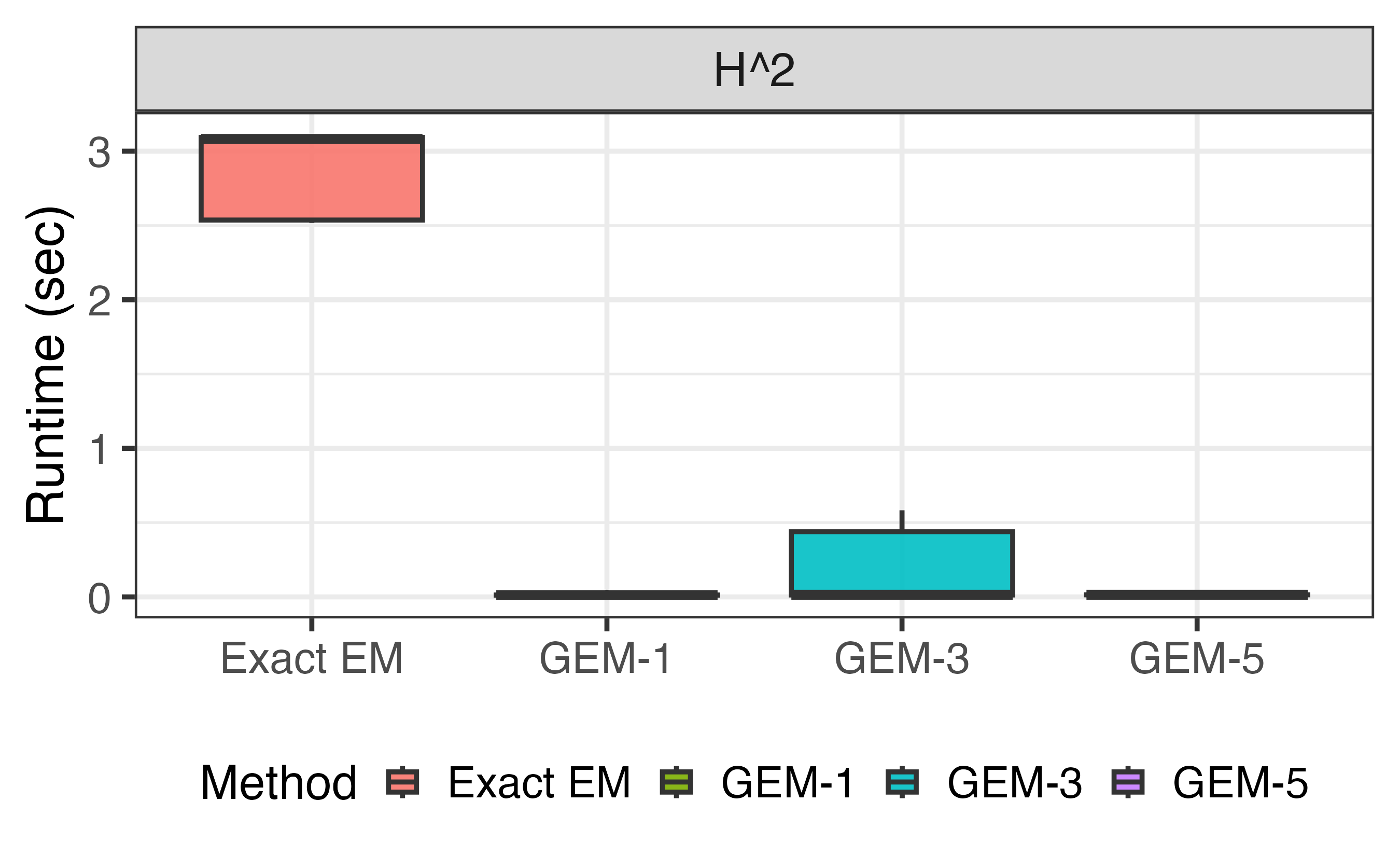}
    \caption{Runtime.}
\end{subfigure}
\hfill
\begin{subfigure}{0.32\textwidth}
    \centering
    \includegraphics[width=\textwidth]{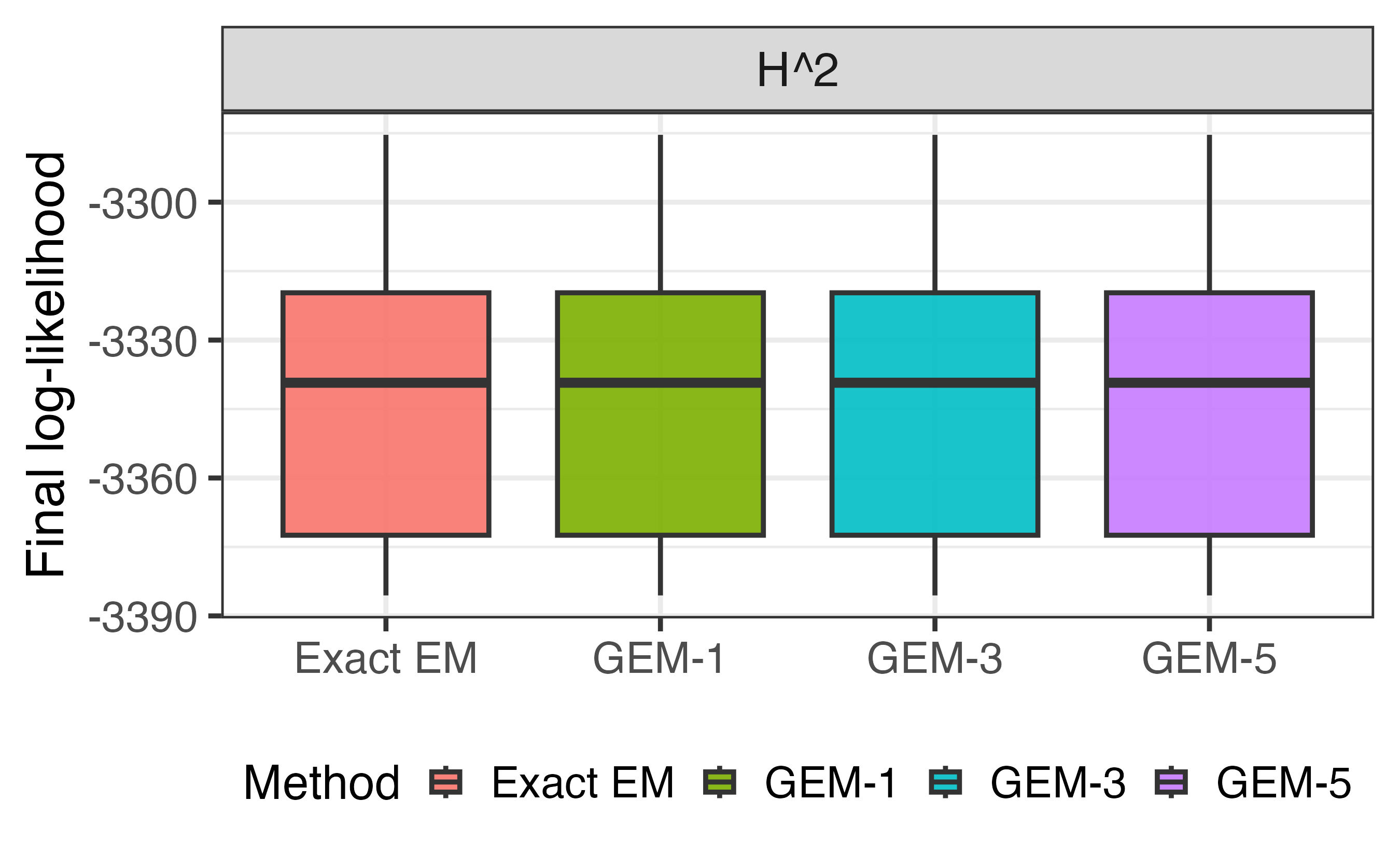}
    \caption{Final log-likelihood.}
\end{subfigure}
\hfill
\begin{subfigure}{0.32\textwidth}
    \centering
    \includegraphics[width=\textwidth]{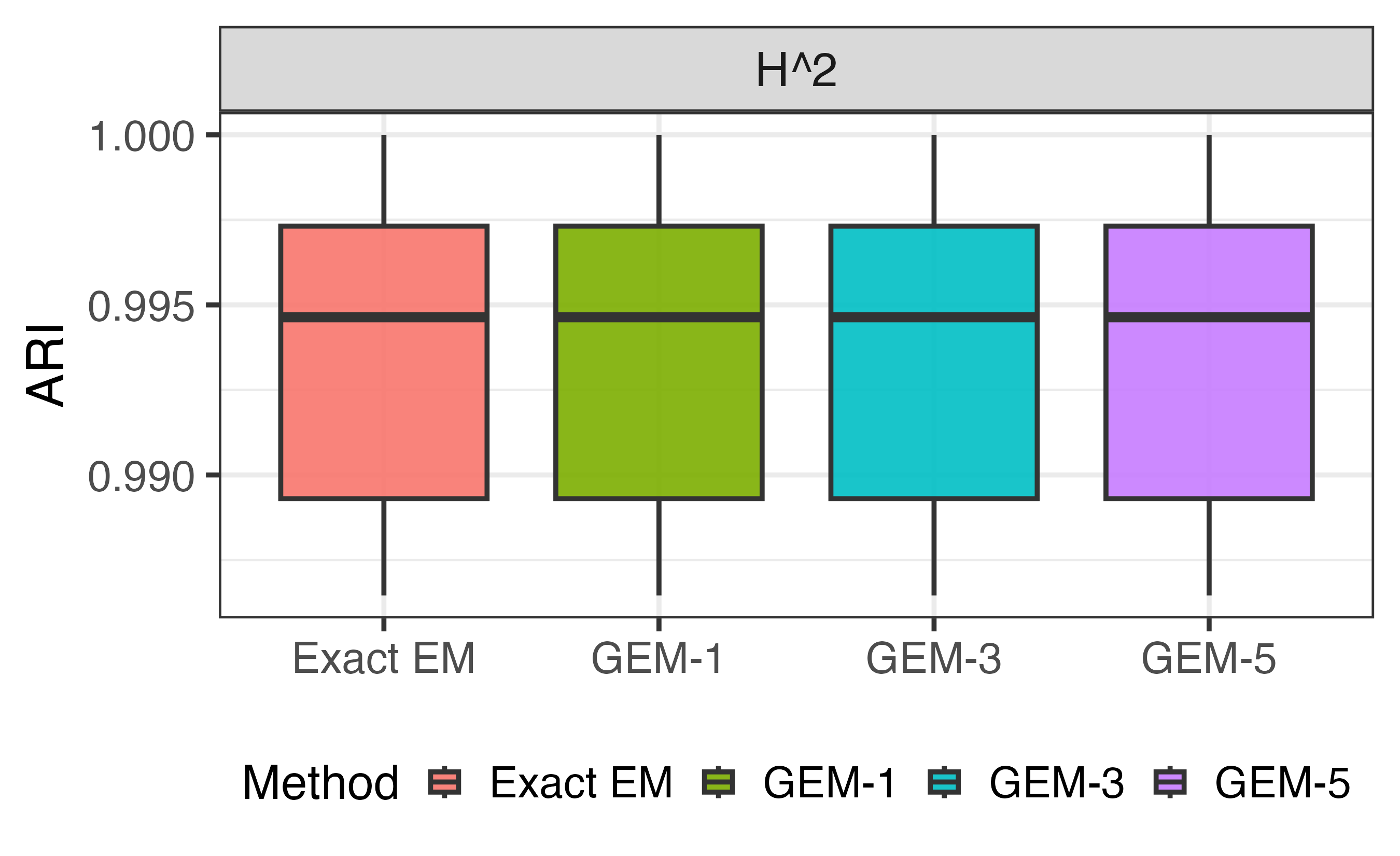}
    \caption{Adjusted Rand index.}
\end{subfigure}
\caption{
Endpoint comparison of exact EM and generalized EM variants. 
Boxplots show runtime, final observed log-likelihood, and adjusted Rand index across Monte Carlo replicates.
}
\label{fig:sim2-emgem-boxplots}
\end{figure}

\begin{figure}[t]
\centering
\includegraphics[width=0.62\textwidth]{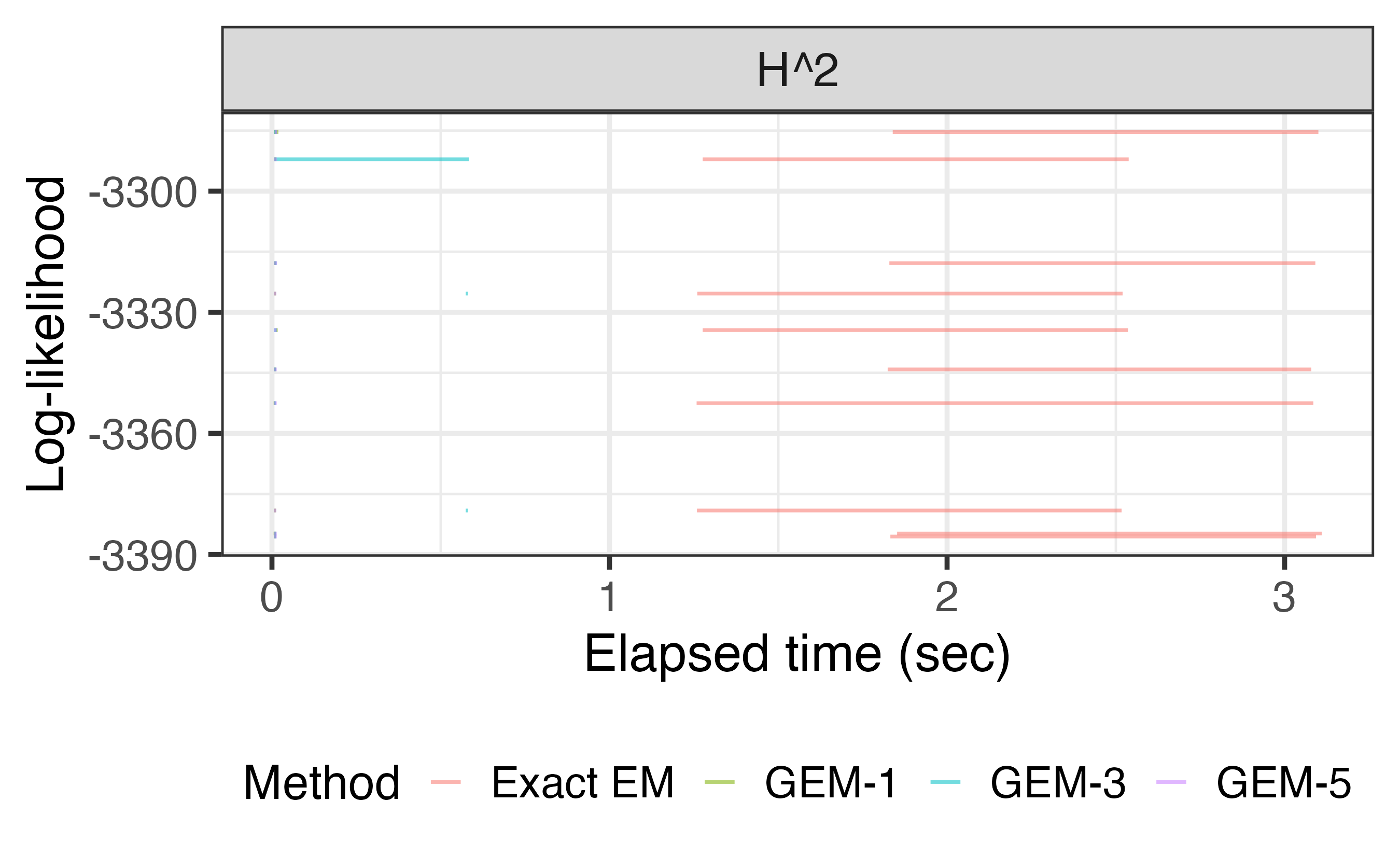}
\caption{
Observed log-likelihood trajectories over elapsed time for exact EM and generalized EM variants. 
Each curve records the likelihood progression during fitting.
}\label{fig:sim2-emgem-trace}
\end{figure}

\paragraph{Exact EM versus GEM.}
Finally, we compare exact EM with three generalized EM variants that use \(L=1\), \(L=3\), and \(L=5\) inner MM updates per component in the location step. All methods are initialized identically within each replicate so that the comparison isolates the optimization routine rather than the effect of initialization. In this experiment we fix \(n=1000\) and use the true value \(K=4\). Table~\ref{tab:sim2-emgem-main} reports the resulting summary statistics, while Figures~\ref{fig:sim2-emgem-boxplots} and \ref{fig:sim2-emgem-trace} display the endpoint distributions and the log-likelihood trajectories over elapsed time.

\begin{table}[t]
\centering
\small
\caption{
Comparison of exact EM and generalized EM variants for the four-component mixture model. 
GEM-\(L\) denotes generalized EM with \(L\) inner MM updates per component. 
Reported values are medians with interquartile ranges in parentheses.
}
\label{tab:sim2-emgem-main}
% Simulation 2C: exact EM versus GEM summary.
% tab:sim2-emgem-main
\begin{tabular}{ccccccc}
\hline
Method & Center error & Rel. scale error & ARI & Final logLik & Runtime (sec) & Inner steps \\
\hline
Exact EM & 0.041 (0.006) & 0.057 (0.020) & 0.995 (0.008) & -3339.28 (52.74) & 3.082 (0.556) &  2400.0 (0.0) \\
GEM-1 & 0.041 (0.006) & 0.057 (0.020) & 0.995 (0.008) & -3339.28 (52.74) & 0.012 (0.001) &  8.0 (0.0) \\
GEM-3 & 0.041 (0.006) & 0.057 (0.020) & 0.995 (0.008) & -3339.28 (52.74) & 0.014 (0.425) &  24.0 (0.0) \\
GEM-5 & 0.041 (0.006) & 0.057 (0.020) & 0.995 (0.008) & -3339.28 (52.74) & 0.014 (0.000) &  40.0 (0.0) \\
\hline
\end{tabular}

\end{table}

The statistical outputs are essentially indistinguishable across the four methods. In median, all procedures achieve the same average center error \(0.041\), the same relative scale error \(0.057\), the same ARI \(0.995\), and the same final log-likelihood \(-3339.28\), up to the reported numerical precision. Moreover, all methods converge successfully in every replicate, and all require exactly two outer iterations in this well-separated setting. The substantive difference is therefore computational. Exact EM has median runtime \(3.082\) seconds and uses \(2400\) total inner MM steps, whereas GEM-1, GEM-3, and GEM-5 reduce the median runtime to \(0.012\), \(0.014\), and \(0.014\) seconds, respectively, while requiring only \(8\), \(24\), and \(40\) inner MM steps. Thus, in the present setting, the computational gain is dramatic and comes almost entirely from replacing exact barycenter solves by a small number of truncated MM updates. Figure~\ref{fig:sim2-emgem-trace} makes this especially clear: the GEM variants reach essentially the same log-likelihood level as exact EM in a negligible fraction of the time. Although GEM-3 shows some variability in runtime across replicates, its computational cost remains far below that of exact EM.

Taken together, the three parts of Simulation~2 provide a coherent picture of the proposed methodology. When the model is correctly specified, the Riemannian Gaussian mixture estimator accurately recovers component parameters and cluster assignments. Standard information criteria behave in a familiar and interpretable manner, with BIC giving particularly reliable recovery of the true number of components. Finally, the GEM variants achieve the same statistical performance and final likelihood as exact EM while reducing the computational cost by orders of magnitude. In this sense, the hyperbolic mixture model inherits the favorable inferential behavior of classical mixture methodology while benefiting from a substantial algorithmic speedup through truncated inner optimization.

\subsection{Real data examples}\label{subsec:realdata-networks}

We next apply the proposed methodology to two network datasets embedded in the low-dimensional hyperbolic space.  The purpose of this experiment is to examine how finite Riemannian Gaussian mixtures behave on real hyperbolic embeddings, where the data are not generated from the model and where external labels need not coincide with density-based mixture components. We consider Zachary's karate club network and the UK faculty friendship network. The karate network has a well-known binary club split, while the UK faculty network contains a vertex-level group attribute and exhibits a more diffuse social structure.

For each network, we first computed a two-dimensional hyperbolic embedding using the Hydra method \citep{keller-ressel_2020_HydraMethodStrainminimizing} applied to graph distances among nodes obtained from the shortest path problem \cite{dijkstra_1959_NoteTwoProblems}. The resulting coordinates were converted to the hyperboloid convention used throughout this paper. The fitted mixture model was estimated in the hyperboloid coordinates, while the Poincar\'e disk representation was used only for visualization. For the karate network, we considered \(K=2,\ldots,6\), and for the UK faculty network, we considered \(K=2,\ldots,8\). For each fitted model, we computed AIC, BIC, and HQIC, and used the BIC-selected model for subsequent interpretation. External vertex labels were not used during estimation; they were used only for post hoc comparison through adjusted Rand index (ARI), normalized mutual information (NMI), and purity.

\begin{figure}[t]
\centering
\includegraphics[width=0.7\textwidth]{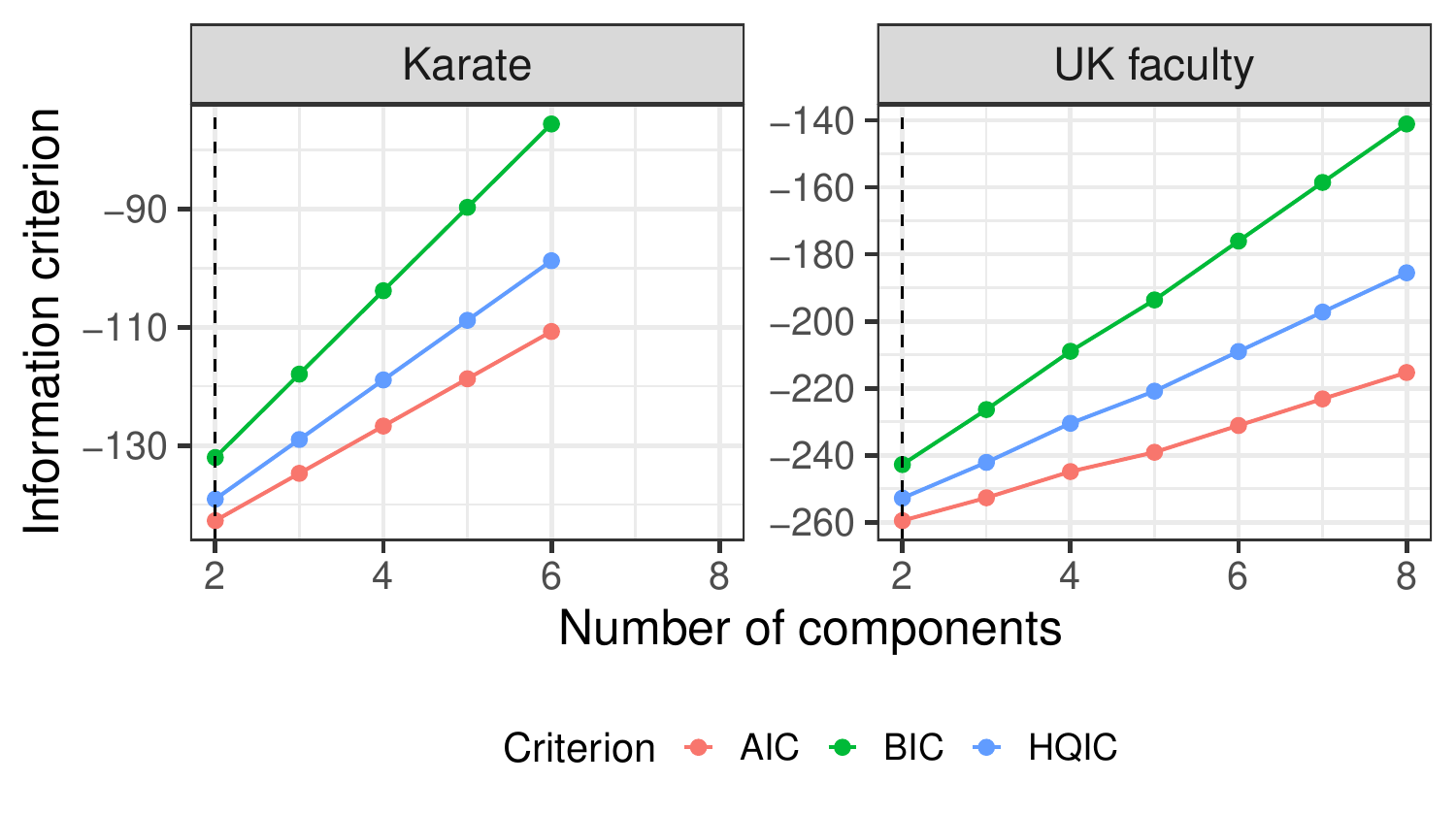}
\caption{
Information criteria for the real network examples. 
Curves show AIC, BIC, and HQIC over candidate numbers of components. The dashed vertical line marks the BIC-selected value.
}
\label{fig:realdata-network-ic}
\end{figure}

Table~\ref{tab:realdata-network-main} summarizes the selected models and their agreement with the external labels. Table~\ref{tab:realdata-network-components} gives component-level summaries for the BIC-selected mixtures. Figure~\ref{fig:realdata-network-ic} shows the information criteria as functions of \(K\), and Figure~\ref{fig:realdata-network} compares the external labels with the mixture-based assignments in the Poincar\'e disk.

\begin{table}[t]
\centering
\small
\caption{
Real network examples. 
The mixture model is fitted to Hydra hyperbolic embeddings, and the BIC-selected model is reported. 
External labels are used only for post hoc comparison.
}
\label{tab:realdata-network-main}
% Network real-data summary.
% tab:realdata-network-main
\begin{tabular}{ccccccccc}
\hline
Dataset & $n$ & Label source & $K$  & ARI & NMI & Purity & Mean max resp. & Mean entropy \\
\hline
Karate & 34 & Club split & 2   & 0.000 & 0.000 & 0.529 & 0.786 & 0.750 \\
UK faculty & 81 & Group  & 2 & 0.007 & 0.040 & 0.420 & 0.977 & 0.131 \\
\hline
\end{tabular}

\end{table}

\begin{table}[t]
\centering
\small
\caption{
Component summaries for the BIC-selected Riemannian Gaussian mixtures fitted to the real network embeddings.
}\label{tab:realdata-network-components}
% Component summaries for BIC-selected models.
% tab:realdata-network-components
\begin{tabular}{llccccc}
\hline
Dataset & Component & pi & beta & sigma & Assigned size & Assigned prop. \\
\hline
Karate & 1 & 0.214 & 50.000 & 0.100 &  0 & 0.000 \\
Karate & 2 & 0.786 & 50.000 & 0.100 & 34 & 1.000 \\
UK faculty & 1 & 0.954 & 50.000 & 0.100 & 79 & 0.975 \\
UK faculty & 2 & 0.046 & 18.782 & 0.163 &  2 & 0.025 \\
\hline
\end{tabular}

\end{table}

\begin{figure}[t]
\centering
\includegraphics[width=\textwidth]{real_report/fig_realdata_combined.png}
\caption{
Hyperbolic embeddings for two network datasets in the Poincar\'e disk model. 
Nodes are colored by (a) external labels and (b) mixture-based cluster assignments under the BIC-selected model.
}
\label{fig:realdata-network}
\end{figure}

For both datasets, the information criteria select a two-component model. In the karate network, AIC, BIC, and HQIC all choose \(K=2\), which is consistent with the binary nature of the classical club split. However, the mixture components do not reproduce the external club labels under hard maximum-posterior assignment. The ARI and NMI are both \(0.000\), and the purity is \(0.529\). This behavior is also visible in Figure~\ref{fig:realdata-network}: the external labels form the familiar two-group annotation, whereas the fitted Riemannian Gaussian mixture represents the embedded points through a soft two-component density. Table~\ref{tab:realdata-network-components} shows that the estimated mixing weights are \(0.214\) and \(0.786\), but all hard assignments fall into the second component. Thus, in this example, the two fitted density components are overlapping rather than producing a hard partition aligned with the club split.

The UK faculty network exhibits a different but related pattern. Again, all three information criteria select \(K=2\), while the external group attribute contains four groups. The BIC-selected mixture therefore gives a coarser geometric summary than the external labeling. The ARI is \(0.007\), the NMI is \(0.040\), and the purity is \(0.420\), indicating weak agreement with the provided group labels. The component summary shows that one component dominates the embedding, with \(79\) of \(81\) vertices assigned to it under hard classification, while the second component captures a small pair of peripheral vertices. Thus, for the UK faculty data, the fitted model identifies a dominant central density region together with a small secondary component, rather than recovering the four external groups.

These examples illustrate the role of the proposed method as a likelihood-based density modeling tool for hyperbolic embeddings.  The fitted components are determined by the geometry and density structure of the embedded points, not by external metadata or community labels.  Consequently, agreement with external labels is not guaranteed, especially when those labels describe a different organizational feature of the network.  The karate example shows a soft overlapping mixture on a small benchmark network, while the UK faculty example gives a parsimonious two-component summary of a more diffuse social network.

%%%%%%%%%%%%%%%%%%%%%%%%%%%%%%%%%%%%%%%%%%%%%%%%%%%%%%%%%%%%%%%%%%%%%%%%%%%%%%%%%
%%%%%%%%%%%%%%%%%%%%%%%%%%%%%%%%%%%%%%%%%%%%%%%%%%%%%%%%%%%%%%%%%%%%%%%%%%%%%%%%%
\section{Conclusion}\label{sec:conclusion}

This paper developed a likelihood-based framework for finite mixture modeling with Riemannian Gaussian distributions on hyperbolic space. The central idea was to treat the weighted single-component likelihood as the basic computational primitive. This is natural because the M-step of a finite mixture model consists of component-wise weighted estimation problems, with posterior responsibilities acting as weights. Under the hyperboloid model, this problem separates cleanly: the location update is a weighted Fr\'echet mean, and the scale update is a one-dimensional convex optimization problem governed by the log-normalizer.

Methodologically, we formulated constrained finite mixtures of Riemannian Gaussian components and derived exact EM and generalized EM algorithms. Exact EM solves each weighted barycenter subproblem to completion, whereas GEM uses a finite number of hyperbolic MM steps to obtain a descent-producing location update. This separation isolates the main computational cost of the M-step and provides a simple way to trade off inner optimization effort against runtime without changing the statistical objective.

Theoretically, we established the main properties needed to support likelihood-based computation. For the single-component model, we proved existence and uniqueness of the weighted location estimator and characterized the scale estimator through a radial moment-matching equation. For finite mixtures, we showed that the unrestricted likelihood is singular, introduced a compact constrained parameter space, and proved existence of the constrained maximum likelihood estimator. We also proved monotonicity of exact EM, fixed-point convergence of accumulation points under an interior compactness condition, and likelihood ascent for the GEM scheme based on truncated MM barycenter updates.

The numerical experiments supported the statistical and computational development. The weighted single-component simulations showed stable and accurate estimation under heterogeneous observation weights. The four-component mixture simulations demonstrated accurate parameter recovery, reliable model selection by BIC, and substantial computational savings from GEM relative to exact EM. The real network examples illustrated how the method can be applied to hyperbolic embeddings of graph data, while also showing that density-based mixture components need not coincide with external metadata labels.

Several extensions remain natural. The present work focuses on isotropic Riemannian Gaussian components; anisotropic or structured-covariance analogues would allow richer local geometry but require more involved normalization and computation. The curvature of the hyperbolic space is fixed at \(-1\), and incorporating curvature estimation into the mixture model would connect the present work to recent developments in statistical hyperbolic latent-space modeling. Finally, the weighted formulation developed here provides a natural starting point for penalized mixtures, classification EM, and Bayesian or nonparametric extensions on hyperbolic space.

Overall, the proposed framework provides a statistical and computational foundation for model-based clustering and density estimation on hyperbolic space. By combining intrinsic likelihood modeling, explicit hyperboloid computations, EM-type algorithms, and supporting theory, it offers a practical route for analyzing data whose natural geometry is non-Euclidean and negatively curved.

\bibliographystyle{dcu}
\bibliography{references}

%%%%%%%%%%%%%%%%%%%%%%%%%%%%%%%%%%%%%%%%%%%%%%%%%%%%%%%%%%%%%%%%%%%%%%%%%%%%%%%%%
%\newpage 
%\appendix
%this is appendix
%\section*{Supplementary Material}

\end{document}